\documentclass[
preprint,
 amsmath,amssymb,
 aps,
]{revtex4-2}

\usepackage{graphicx}% Include figure files
\usepackage{wrapft}
\usepackage{dcolumn}% Align table columns on decimal point
\usepackage{amscd,amsmath,theorem,amssymb}
\usepackage{slashed}
\usepackage{bm}
\usepackage{amsfonts} 
\usepackage[normalem]{ulem}

\usepackage{url}

\newcommand{\be}{\begin{equation}}
\newcommand{\ee}{\end{equation}}
\newcommand{\ba}{\begin{align}}
\newcommand{\ea}{\end{align}}

\begin{document}

\title{Holographic quantum singularity} 
\author{Izumi Tanaka}
 \email{escargoviolin@infoseek.jp} 
\affiliation{%
 Yotuya-gakuin, Fukuoka 810-0041 Japan 
}%

%\catchline{}{}{}{}{}
%
%%%%%%%%%%%%%%%%%%%%%%%%%%%%%%%%%%%%%%%%%%%%%%%%%%%%%%%%%%%%%%%%%%%%

%\title{Holographic quantum singularity
%\footnote{For the title, try not to 
%use more than 3 lines. Typeset in 10 pt roman, uppercase and boldface.}
%}

%\author{Izumi Tanaka%\footnote{
%Typeset names in 8 pt roman, uppercase. Use footnote to indicate 
%permanent address of author.}
%}

%\address{Yotsuya, Gakuin\\
%Fukuoka, 810-0041, Japan\\
%\,\footnote{State completely without 
%abbreviations, the affiliation and mailing address, including country. 
%Typeset in 8 pt italic.
%escargoviolin@infoseek.jp}
%\footnote{Typeset author's e-mail 
%&address in 8pt italic}

%\begin{history}
%\received{(Day Month Year)}
%\revised{(Day Month Year)}
%\end{history}

\begin{abstract}
In this study, we addressed the influence of quantum singularity on the topological state. 
The quantum singularity creates the defect in the momentum space ubiquitously and leads to the phase transition for the topological material. The kinetic equation reveals that the defect generates an anomaly without the characteristic energy scale.

In the holographic model, the three-dimensional dislocations map into the gravitational bulk as domain walls extending along the AdS radial direction from the boundary.
The creation/annihilation of the domain wall causes the quantum phase transition by 't Hooft anomaly generation and is controlled by the gauge field. In other words, the phase transition is realized by the anomaly inflow. 
 %%%%%
 
This 't Hooft anomaly is caused by a phase ambiguity of the ground state resulting from the singularity in parameter space. 
This singularity gives the basis for the boundary's topological state with the Berry connection. 
't Hooft anomaly's renormalization group invariance shows that the total Berry flux is conserved in the UV layer to the IR layer.

Phase transition entails domain wall constitution, which generates the entropy from 
the non-universal form or quantum entropy correction.

\end{abstract}

%\keywords{Quantum singularity; Holographic principle; 't Hooft anomaly.}

\maketitle

\section{Introduction}
\label{Intro}
%%%%%%%%%%%%%%%%%%%%%%%%%%%
%%%%%%%%%%%%%%%%%%%%%%%%%%%
%%%%%%%%%%%%%%%%%%%%%%%%%%%
The low energy theory of a strongly coupled fermionic system provides the universality class characterized by high energy physics's symmetry. 
The universal class is determined by topologically 
stable Fermi points in momentum space. 
From the topological viewpoint, the topological matter and the quantum vacuum of the Standard Model share the common universal nature.
Non-trivial topological invariants classify gapless semimetal states and fully gapped states. 
The topological quantum phase transition occurs between different topological number states and characterizes the fundamental process for the topological substance \cite{Volovik_SM}.

In a strongly coupled system, the applicability of a single wave function is not clear. 
Besides, the fermionic quasiparticle picture and relevant topology in momentum space are also 
unclear. 
A scheme to understand the strongly coupled system utilizes the fact that 
the low-energy asymptote of the Green's function in a strongly coupled system gives the symmetry and topology information.  
Apart from this, these days, 
the holographic principle had shed light on a strongly coupled condensed matter system. 

The application of the holographic principle \cite{Maldacena} to condensed matter physics has been in the spotlight 
 after the discovery of holographic superconductor models \cite{HHH2008,Gubser-Pufu2008,Roberts-Hartnoll2008,Hutasoit-Siopsis-Therrien2014}. 
This principle has also been applied to holographic models of semimetal \cite{Landsteiner-Liu2015, Landsteiner2015, Liu-Sun2018TI}. 
From this progress, the gravity dual phenological model presently serves as a guiding principle 
that will help us understand the topological substance in a strongly coupled system where perturbative methods are no longer available.

Quantum singularity affects the character of topological substances. 
Historically, general relativity establishes the concept of singularities in spacetime. 
In general relativity, geodesic completeness defines a singular spacetime. 
Spacetime's geodesically incompleteness represents some test particle's evolution that is not defined after a finite proper time. 
Concretely, the classical singularity shows the sudden end of a classical particle path. 
On the other hand, quantum completeness or quantum-mechanically non-singular represents a unique unitary time evolution for test fields propagating on an underlying background. 

Quantum completeness renders a spatial part of the wave operator to be essentially self-adjoint. Geodesically complete Riemannian manifolds give the corresponding Laplacian essentially self-adjoint, i.e., classical completeness assures quantum completeness  \cite{Chernoff1973}.
However, there is geodesically incomplete static spacetime being quantum complete \cite{Horowitz-Marolf1995}.
From this viewpoint, the quantum singularity is where the wave operator's spatial part is not essentially self-adjoint.  
Moreover, quantum singularity depends on the type of field.

For static globally hyperbolic spacetime, we can define a consistent quantum theory for a single relativistic particle: each one-particle state's energy is identical to that of the corresponding classical particle \cite{Astekar-Magnon1975,Hofmann-Schneider2015}. Meanwhile, the creation and the annihilation of particles occur in the quantum theory of the general time-dependent spacetime. Only quantum field theory (QFT) adequately describes this situation. 
This fact shows that general time-dependent spacetime does not allow the consistent quantum theory of a single particle. 

The modes for Dirac fields in dislocated media can be classified into two groups. First is quantum non-singular modes, which correspond to consistent quantum theory, and second is quantum singular modes, which correspond to the topologically protected defect of the momentum space. The defect of the momentum space is the singularity of Green's function. 

This defect prohibits the formation of the topological number based upon the base space's homotopic character. At the same time, 
the defect permits a new topological number formation around itself. 
 Further, fine-tuned magnetic flux adjusts the effect of the defect. From this property, we can control the phase transition \cite{IT2019}. 

The defect winding number determines the stability of the fermion zero-modes. 
Thus, the fermion zero-modes around the defect are preserved even when particles' interaction is introduced, where the one-particle Hamiltonian is no longer valid. 
Of course, the defect is not affected by the quantum non-singular modes. From this viewpoint, the quantum singularity can be imported to the context of QFT.

The holographic principle is a duality between a $d$ dimensional QFT and $d+1$ dimensional
gravitational theory \cite{Maldacena}.  
Through this principle, we can understand the quantum physics of a strongly coupled many-body system from the classical dynamics of gravity. 

QFT is ideally sliced by the family of trajectories of the renormalization group (RG) flows, 
and the energy scale is an additional coordinate for the QFT. 
As a result, in the holographic principle, $AdS_{d+1}$ gives geometrization of quantum dynamics with renormalization group encoding. 
The lattice version of multiscale entanglement renormalization ansatz (MERA) captures the essence 
of the holographic principle \cite{Swingle2012}. 
The continuous version of MERA has been developed by applying the entanglement renormalization 
to QFTs and gives the holographic metric in an extra dimension \cite{Nozaki-Ryu-Takayanagi2012}.
%%%%%%%%%%%
%%%%%%%%%%%

Utilizing this principle as a stepping stone for considering strongly coupled problems 
leads to revealing new physics.   
The holographic model of condensed matter in zero temperature is initiated in a simple form, where dual conformal field theory (CFT) is defined in boundary Minkowski spacetime, for $AdS_{d+1}$ geometry is a family of copies of Minkowski spaces by the radial coordinate parameterization.    

However, for actual physical systems, the holographic model has to deal with the irregularity in condensed matter by various methods \cite{ZLSS2015}. 
Effects of the disorder on the holographic model is introduced by random chemical potential 
on the boundary \cite{AFZLS2014}. 
Breaking translational symmetry is incorporated from several viewpoints. 
Spatially anisotropic holographic model is discussed from the dilaton field \cite{Koga-Maeda-Tomoda2014}. 
Bianchi spacetime is homogeneous but anisotropic and provides bulk geometries that describe a boundary with 
broken symmetry \cite{Donos-Hartnoll2013,Donos-Gauntlett-Pantelidou2015}. 
Breaking of global translation in the boundary is considered by massive gravity \cite{RGT2011}. 
Spontaneous breaking of translational symmetry is considered by introducing the Chern-Simons term 
\cite{NOP2010,Ooguri-Park2010}.
Holographic impurity models incorporate such as Anderson and Kondo impurity where 
brain intersections describe the physics of defects \cite{HKT2012,EHOJ2013,KKY2010}.  

The screw dislocation is geometrically described as distributional torsion and
gives rise to breaking translational invariant. 
For setting up the holographic quantum singularity, we take the quantum characteristic deeply related to its geometry.   
Dislocation has a similar construction as a hole threading a magnetic flux \cite{Shen2012}. 
Vortices have a connection with magnetic flux, and the holographic model of the two-dimensional vortex has been constructed \cite{Chesler-Hong-Adams}. 
In this model, vortices map into the gravitational bulk as flux tubes extending along the AdS radial direction from the boundary. This result suggests three-dimensional dislocations map into the gravitational bulk as domain walls extending along the AdS radial direction from the boundary. 
In this mapping, we might be careful that the dislocation is a quantum singularity with a specific quantum effect. 
This information leads to the bulk spacetime with dislocated boundary is an adequate option 
to discuss the holographic model of quantum singularity.

This paper aims to consider the effect of quantum singularity on the topological state. To consider the quantum singularity in the topological state, we assume static dislocated spacetime.  
Further, we consider the quantum singularity from the holographic model. 
The organization of the paper is as follows. In Sec. 2, we summarize the phase diagram on dislocated media and study quantum singularity from the kinetic equation.  
In Sec. 3, we study it in the holographic model.
We conclude with some discussions in Sec. 4. 

%%%%%%%%%%%%%%%%%%%%%%%%%%%%%%%
%%%%%%%%%%%%%%%%%%%%%%%%%%%%%%%
%%%%%%%%%%%%%%%%%%%%%%%%%%%%%%%
\section{Quantum singularity in topological state}
%%%%%%%%%%%%%%%%%%%%%%%%%%%%%%%
%%%%%%%%%%%%%%%%%%%%%%%%%%%%%%%
%%%%%%%%%%%%%%%%%%%%%%%%%%%%%%%
%%%%%%%%%%%%%%%%%%%%%%%%%%%%%%%
%%%%%%%%%%%%%%%%%%%%%%%%%%%%%%%
\subsection{Phase diagram of topological state on dislocated media}
%%%%%%%%%%%%%%%%%%%%%%%%%%%%%%%
%%%%%%%%%%%%%%%%%%%%%%%%%%%%%%%
%%%%%%%%%%%%%%%%%%%%%%%%%%%%%%% 
Dislocated media is expressed as the following metric 
\cite{Galtsov-Letelier,Furtado-Berezza-Moraes2001}:
\begin{equation}\label{GLT001}
ds^2=-dt^2+d\rho^2+\rho^2d\varphi^2 +(\gamma_B d\varphi +dz)^2, 
\end{equation}
where $2\pi\gamma_B$ 
is analogous to the Burgers vector.   
Modified Dirac system for the dislocated media is given as follows:
\be\label{Dirac}
\mathcal{L}=\bar{\Psi}\Big( i\gamma^{\mu}\partial_{\mu}+i\gamma^{\mu}\Gamma_{\mu}-M  \Big) \Psi,
\ee 
where $M=M_0+M_2(p_x^2+p_y^2+p_z^2)$.  
State $M_2=0$ corresponds to a trivial topological insulator, and further when $M_0=0$ corresponds 
to gapless semimetal.  
This system presents quantum singularity, which generates a kind of defect in momentum space.

The phase diagram without the dislocation has already known and shown as Fig. {\ref{fig:PhaseT}}(A). 
This phase diagram features the vacuum of the Standard Model \cite{Volovik_SM}. 
As shown in Fig. {\ref{fig:PhaseT}}(B),  
the quantum singularity sets off the phase transition. 
In Fig. {\ref{fig:PhaseT}}, $N$ is the topological invariant, a winding number of the momentum space, and
$\tilde{N}$ is the topological invariant, a winding number of the momentum space around the defect
(See Appendix). The defect originated in quantum singularity induces a quantum phase transition among  
Fig. \ref{fig:PhaseT}(A)  and  Fig. \ref{fig:PhaseT}(B).  
We can control this transition by the magnetic field.
From the phase diagram, the quantum singularity is ubiquitous 
when we consider fermions on dislocated media.

The state 
$M_0=M_2=0 $ in the phase diagram Fig. \ref{fig:PhaseT} also corresponds to Weyl semimetal. 
This Weyl fermion is given by Lorentz breaking Dirac system in curved spacetime \cite{Chen-Wang-Su2006}:
\be\label{c_Dirac}
\mathcal{L}=\bar{\Psi}\Big( i\gamma^{\mu}\partial_{\mu}+i\gamma^{\mu}\Gamma_{\mu}-b_{\mu}e_{(a)}^{\mu}\gamma_5 \gamma^{(a)}-M  \Big) \Psi.
\ee
For simplicity, we assume that the vector $\mathbf{b}$ only has $z$-component $b$, and $M$ is the Dirac field's mass.  

%%%%%%%%
%%%%%%%%
\begin{figure}[htbp]
  \begin{center}
   \includegraphics[width=140mm]{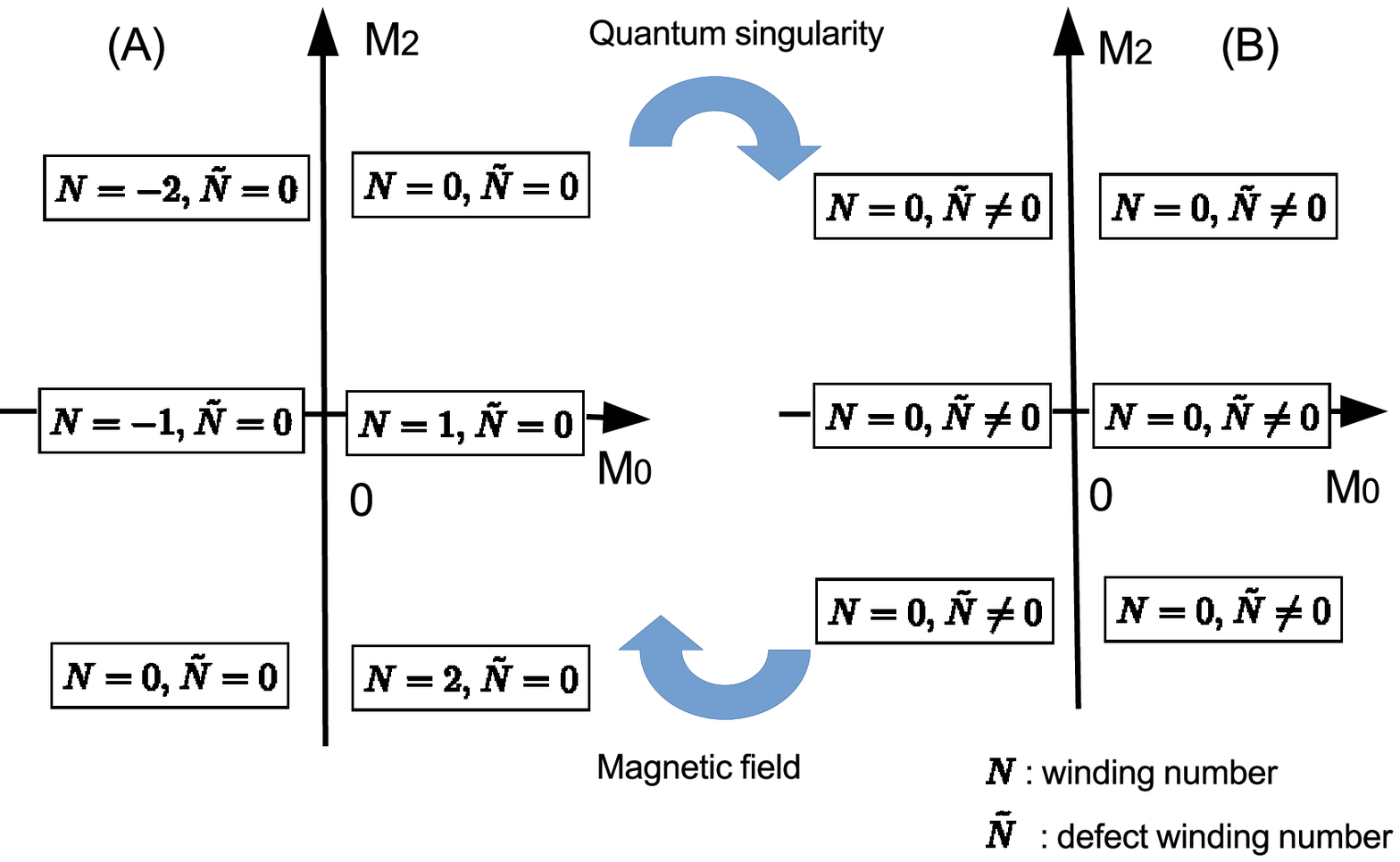}
  \end{center}
  \caption{Phase diagram of topological states of the fermionic field with/without quantum singularity}
%\footnotemark[1]
  \label{fig:PhaseT}
\end{figure} 
%\footnotetext[1]{
%%%%%%%%%%
%%%%%%%%%%

We use the chiral representation of the $\gamma$ matrices 
for Weyl fermion \cite{Pallab-Sumanta2013}. 
\begin{align} 
\gamma^{(0)}= 
\left(    \begin{array}{cc}
0 & 1  \\ 
1 & 0
\end{array} \right), \hspace{1mm}
\gamma^{j}= 
\left(    \begin{array}{cc}
0 &  \sigma^j  \\ 
-\sigma^j & 0
\end{array} \right)(j=1,2,3), \hspace{1mm}
\gamma_{5}= 
\left(    \begin{array}{cc}
-1 & 0  \\ 
0 & 1
\end{array} \right). 
\end{align}             
In these expressions, $\sigma^1$, $\sigma^2$, and $\sigma^3$ are the usual Pauli 
matrices.

From the following ansatz 
\be
\lvert \psi \rangle =
\begin{pmatrix}
u_1(\rho) \\
u_2(\rho)
\end{pmatrix}e^{-iEt+i\ell \varphi + ikz},
\ee
we have the following equation for $u_i (i=1,2)$: 
\be\label{radial01}
u''_1+\frac{1}{\rho}u_1'+\Big((E-M)^2-(k-b)^2-\frac{(\ell -\gamma_B k)^2}{\rho^2}\Big)u_1=0,  
\ee
\be\label{radial02}
u''_2+\frac{1}{\rho}u_2'+\Big((E-M)^2-(k-b)^2-\frac{(\ell +1 -\gamma_B k)^2}{\rho^2}\Big)u_2=0.  
\ee
From Eq.(\ref{radial01}) and Eq.(\ref{radial02}), the radial solutions are given as Bessel and Neumann functions:
\begin{align}\label{TC00}
u_i(r) = C_1& 
J_{\lvert \tau + i -1 \rvert} (\sqrt{(E-M)^2-(k-b)^2}\rho)  \nonumber \\
+C_2&
N_{\lvert \tau + i -1 \rvert} (\sqrt{(E-M)^2-(k-b)^2}\rho) \hspace{2mm} (i=1,2),
\end{align}
where $\tau =\ell -\gamma_B k$.

The equations (\ref{radial01}) and (\ref{radial02}) show that $\gamma_B$ is included in the effective angular momentum, i.e., $\ell \rightarrow \ell -\gamma_B k$. 
Further, by applying the magnetic flux $\Phi$, 
$\ell -\gamma_B k \rightarrow \ell -\gamma_B k -\nu$, where $\nu = \Phi /(\text{flux quantum})  $ is obtained.  
These relations give the same result shown in Dirac fermion. See appendix for detailed discussion.   
Weyl fermions and Dirac fermions on the dislocated metric hold the effect of the quantum singularity in common. 

We can fine-tune the condition to control the quantum singularity by applying the magnetic flux. 
The come-and-go of  Fig. \ref{fig:PhaseT}(A) between Fig. \ref{fig:PhaseT}(B) is realized.  
We summarize the quantum phase transition as follows:
In Fig. \ref{fig:PhaseT}(a), $M_2=0$ corresponds to the trivial insulators.
$M_0$ marks quantum phase transition among the topological insulating phase. 
The defect originated in quantum singularity induces a quantum phase transition among  
Fig. \ref{fig:PhaseT}(A)  and  Fig. \ref{fig:PhaseT}(B). 
We can control this transition by the magnetic field.
From the phase diagram, the quantum singularity is ubiquitous 
when we consider fermions on dislocated media.

%%%%%%%%%%%%%%%%%%%%%%%%%%%%%%%
\subsection{Kinetic equation for Weyl fermions}
%%%%%%%%%%%%%%%%%%%%%%%%%%%%%%%
%%%%%%%%%%%%%%%%%%%i%%%%%%%%%%%%  
We concentrate on Weyl fermions' quantum singularity from the kinetic viewpoint to understand the momentum space defect. 
We consider the classical dynamics of the finite-density of fermions.  
This approach is quite general and valid for the fermionic system \cite{stephanov-yin2012,stone-dwivedi2013}.  
The following action describes
$3+1$-dimensional positive-energy, positive-helicity Weyl particle:   
\be \label{k-a}
S[\mathbf{x}, \mathbf{p}]=\int dt \Big( \mathbf{a}\cdot \dot{\mathbf{x}}-\phi (\mathbf{x}) +\mathbf{p} \cdot \dot{\mathbf{x}}     -\lvert \mathbf{p}  \rvert - \hat{\mathbf{a}}\cdot \dot{\mathbf{p}} \Big) , 
\ee    
where $\mathbf{a}\cdot \dot{\mathbf{x}}$ is the standard coupling of the Maxwell vector potential to the velocity $ \dot{\mathbf{x}}$
of the charged particle.
The momentum-space gauge field $\hat{\mathbf{a}}$ is the adiabatic Berry connection.
This gauge field is obtained from the $E=+\lvert \mathbf{p} \rvert$
eigenvector of  the Weyl Hamiltonian $H=\mathbf{\sigma} \cdot\mathbf{p}$.
From the action defined by Eq.(\ref{k-a}), 
$6+1$ phase space current $(\rho, \rho \dot{\mathbf{x}}, \rho\dot{\mathbf{p}})$ 
obeys  the continuity equation with source:
\be
\frac{\partial \rho}{\partial t}+\frac{\partial (\rho \dot{\mathbf{x}})}{\partial \mathbf{x}}+ \frac{\partial (\rho \dot{\mathbf{p}})}{\partial \mathbf{p}}=
2\pi \mathbf{E \cdot B \delta(p)}.
\ee
The last term shows the quantum effect, which injects particle number violation into the classical description.  
Integrating over the momentum $\mathbf{p}$, we obtain standard expression of the electromagnetic anomaly at zero temperature. 
\be
\frac{\partial n}{\partial t}+\mathbf{\nabla}\cdot \mathbf{j}=
\frac{1}{4\pi^2}\mathbf{E}\cdot \mathbf{B},
\ee
where $n=\int \rho \frac{d^3\mathbf{p}}{(2\pi)^3}$ and 
$\mathbf{j}=\int \rho \dot{\mathbf{x}} \frac{d^3\mathbf{p}}{(2\pi)^3}$.
In this calculation, the singular point $\mathbf{p}=0$ is excluded.

If we consider the dislocated media described in Eq.(\ref{GLT001}), 
we must take the torsion tensor \cite{Furtado-Berezza-Moraes2001,Ishihara-Mizushima-Tsuruta-Fujimoto2019,Marques-Furtado-Bezerra-Fernando2001}.
We have $\mathbf{e}^a= e^a_{\mu}dx^{\mu}$, specifically, $\mathbf{e}^1=\gamma_B d\varphi +dz$, $\mathbf{e}^2=d\rho$, $\mathbf{e}^3=\rho d\varphi$ for the spatial part of the metric in Eq.(\ref{GLT01}).
Torsion can be expressed as
\be
T^a_{\mu\nu}=\partial_{\mu}e^a_{\nu}- \partial_{\nu}e^a_{\mu}, 
\ee
where the two-form component $\mathbf{T}^a =T^a_{\mu\lambda}dx^{\mu}\wedge dx^{\lambda}$.  
Further, $T^{\nu}_{\mu \lambda}=e_a^{\nu}T^a_{\mu\lambda}$,  where $e_a^{\nu}$ is the inverse of $e^a_{\nu}$. 
The Burgers vector can be viewed as a flux of torsion:
\be
\int \mathbf{T}^1=\oint \mathbf{e}^1=2\pi \gamma_B=b,
\ee 
where $\mathbf{T}^1=2\pi \gamma_B\delta^2(\rho)d\rho \wedge d\phi$. 
Then, Euler-Lagrange equation consideration is given as follows:   
\be
\frac{d}{dt}\Big( \frac{\partial \mathcal{L} }{\partial \dot{x}^{\mu}} \Big) - \frac{\partial \mathcal{L}}{\partial x^{\mu}} =T^{\nu}_{\mu\lambda}
\dot{x}^{\lambda}\frac{\partial \mathcal{L}}{\partial \dot{x}^{\nu}}.
\ee
The torsion generates the following effective magnetic field 
acting on quasiparticle:
\be\label{eff-mag}
\mathfrak{B}=\mathbf{T}^\mu (p_{\mu}+\hat{a}_{\mu}),
\ee where $(\mathbf{T}^\mu)^{\nu}=\frac{1}{2}\varepsilon^{\nu\lambda\rho}T^{\mu}_{\lambda\rho}$ includes the 
two-dimensional delta function.
Integrating the corresponding continuity equation over the momentum $\mathbf{p}$  
does not remove the infinity from the delta-function defined at the coordinate. 
However, from Eq.(\ref{eff-mag}), $\rho=0$, i.e., the quantum singularity, gives the quantum input for this system.

We take advantage of the defect generated from the quantum singularity. From the discussion in \S 2.1, 
the defect's nature shows that the defect can break the Liouville theorem. 
So the phase space measure is not conserved, and the following effective equation can be obtained:
\be
\frac{\partial n}{\partial t}+\mathbf{\nabla}\cdot \mathbf{j}=
\mathcal{I}_{Defect},
\ee
$\mathcal{I}_{Defect}$ shows that anomaly is generated from the defect: 
$\mathcal{T}_{Defect}$ provides a different physics in the original system and leads to the non-conservation of space measure.    
The size of the defect is specified by $\lvert \tau + j - 1\rvert < 1$ ($j=1,2$), where $\tau =\ell -\gamma_Bk -\nu$. 
From this relation, $\frac{\ell -\nu -1}{\gamma_B}  < k < \frac{\ell -\nu +2}{\gamma_B}$, where $k$ is the $z$-component of momentum.
Inside this region, the motions of particles are considered fully quantum mechanically.  
From this viewpoint, $\mathcal{T}_{Defect}$ is a function of  
$2\pi\gamma_B$, which determines the size and the 
location of the defect in the momentum space: $\mathcal{T}_{Defect} = f (\lvert \tau + j - 1\rvert)$. $\mathcal{T}_{Defect}$ expresses the quantum effect in momentum space, which is specified by the dislocation.  

The defect in the momentum space causes the 
anomaly, followed by the phase transition.
The anomaly that originated from the defect is ubiquitous in the phase diagram 
Fig. \ref{fig:PhaseT}.   
The anomaly from the defect occurs without the 
characteristic energy scale as a symmetry breaking. 
Further, this anomaly affects low energy physics.

%%%%%%%%%%%%%%%%%%%%%%%%%%%%%%%
%%%%%%%%%%%%%%%%%%%%%%%%%%%%%%%
%%%%%%%%%%%%%%%%%%%%%%%%%%%%%%%
\section{Holographic quantum singularity}
%%%%%%%%%%%%%%%%%%%%%%%%%%%%%%%
%%%%%%%%%%%%%%%%%%%%%%%%%%%%%%%
%%%%%%%%%%%%%%%%%%%%%%%%%%%%%%%
%%%%%%%%%%%%%%%%%%%%%%%%%%%%%%%
\subsection{Dislocated boundary solutions} 
%%%%%%%%%%%%%%%%%%%%%%%%%%%%%%%
%%%%%%%%%%%%%%%%%%%%%%%%%%%%%%%
In this section, we consider the holographic model of quantum singularity. 
We pay attention to the characteristic realized by the quantum singularity of the topological insulator. 
Because quantum singularity is characterized by Green's function and protected topologically,
symmetry and topology information is inherited in the strongly coupled system. 
Although the gravitational calculation is purely classical in the bulk, this characteristic is captured by 
the classical quantities in the bulk. 

On the analogy of the holographic model of the two-dimensional vortex, three-dimensional dislocations map into the gravitational bulk as domain walls extending along the AdS radial direction from the boundary. The domain wall affects the symmetry of the bulk.

It is appropriate to consider the theory, not on four-dimensional Minkowski spacetime but the dislocated spacetime where translational invariance is broken.  
In other words, the boundary has distributional torsion discussed in the previous section.   
This case is equivalent to considering the theory is deformed by operators breaking translational invariance 
\cite{Donos-Hartnoll2013,Donos-Gauntlett-Pantelidou2015}.
\begin{equation}
ds^2=-dt^2+d\rho^2+\rho^2d\varphi^2 +(\gamma_B d\varphi +dz)^2.
\end{equation} 
The five-dimensional Einstein-Hilbert action given by
\be\label{E-H}
S_0=\int d^5x \sqrt{-g} (R+12),
\ee
where we have set $16\pi G =1$ and  
the cosmological constant to be $\Lambda =-6$. 
The equation of  motion is given by 
\be\label{EOM}
R_{\mu\nu}=-4g_{\mu\nu}.
\ee
The metric ansatz for the  
solution is given by 
\be
ds^2=-G(r)F^2(r)dt^2+G(r)^{-1}dr^2+H^2(r)d\rho^2+r^2\Big( \ \rho^2d\varphi +(\gamma_B d\varphi +dz)^2  \Big),
\ee
where $F(r)$, $G(r)$, $H(r)$ are functions of the radial coordinate $r$.
Substituting the ansatz into Eq.(\ref{EOM}), the following systems of equations are obtained:
\be
\frac{1}{2rH} \Big(  rFG''H+2rF''GH+rFG'H'+3rF'G'H+2rF'GH'+2FG'H+4F'GH     \Big)  = 4F, 
\ee
\be
\frac{1}{2rFH}\Big(   rFG''H+2rF''GH+2rFGH''+rFG'H'+3rF'G'H+2FG'H  \Big)= 4, 
\ee
\be
\frac{-rH'+H}{r\rho H}=0, 
\ee
\be
\frac{1}{rF}\Big(  rFGH''+rFG'H'+rF'GH'+2FGH'   \Big) = 4H, 
\ee
%%%
\be
\frac{1}{FH}\Big(   rFG'H+rFGH'+rF'GH+FGH   \Big)= 4r^2. 
\ee
From these equations, we have the following: 
\be
F=Const,\hspace{1mm} G=r^2- \frac{Const}{r^2},\hspace{1mm}  H=r.
\ee 
The above conditions give zero temperature or finite temperature solutions:
\be 
\label{zero-tem}
ds^2=r^2\{ -dt^2 +d\rho^2 +\rho^2d\varphi^2+(\gamma_Bd\varphi +dz)^2  \}+\frac{1}{r^2}dr^2,
\ee
or 
\be 
ds^2=-(r^2-\frac{r_+}{r^2})dt^2 +\frac{dr^2}{r^2-\frac{r_+}{r^2}}+r^2\{   d\rho^2 +\rho^2d\varphi^2+(\gamma_Bd\varphi +dz)^2  \}.
\ee
To understand the ground state of the system, we 
devote our attention to zero temperature solutions.

%%%%%%%%%%%%%%%%%%%%%
%%%%%%%%%%%%%%%%%%%%%
%%%%%%%%%%%%%%%%%%%%%
\subsection{Holographic fermions} 
%%%%%%%%%%%%%%%%%%%%%
%%%%%%%%%%%%%%%%%%%%%
In Fig. \ref{fig:PhaseT}, the defect that originated from the quantum singularity is ubiquitous 
in topological matters. 
The origin of Fig. \ref{fig:PhaseT}, i.e., the massless Dirac field, is the case.    
Because massless Dirac excitation is generated for pure AdS case, 
we consider the corresponding candidate in dislocated boundary solution. 
In five-dimensional spacetime, the bulk four-component spinor corresponds to a two-component spinor of the dual field theory in four dimensions. 
Selecting two-spinors with opposite mass and axial charge in the bulk and choose one spinor with standard quantization while the other spinor with alternative quantization, we have a four-component spinor with opposite chiralities \cite{ Liu-Sun2018TI,Iqbal:2009fd}.

The effect of the quantum singularity 
is also considered ubiquitous in the strongly coupled fermionic system. 
%%%%%%%%%%%%%
The dual holographic description has the following action:
\begin{align}\label{action} 
S&=S_0+S_1  \nonumber \\ 
S_1= \int d^5x \sqrt{-g} \Big(& -\frac{1}{4}F^{\mu\nu}F_{\mu\nu}+ 
\bar{\Psi}_1 \big( i\gamma^\mu \partial_{\mu} +i\gamma^{\mu}\Gamma_{\mu}-m+A_{\mu}\gamma^{\mu}  \big)\Psi_1 \nonumber \\
&+\bar{\Psi}_2 \big( i\gamma^\mu \partial_{\mu} +i\gamma^{\mu}\Gamma_{\mu}+m-A_{\mu}\gamma^{\mu}  \big)\Psi_2  
 \Big) +S_{boundary}
\end{align}

$S_{boundary}$ ensures that the total action has a well defined variational principle. 
We use the following expressions \cite{Oliveria-Tiomno1962}. 
\be
\gamma^{\mu}\Gamma_{\mu}=\gamma^{a}\hat{\Gamma}_{(a)}+\check{\Gamma}, \hspace{1mm}  \hat{\Gamma}_{(a)}=\frac{1}{2}(\partial_{\mu}e^{\nu}_a+e^{\rho}_a\Gamma^{\mu}_{\rho \mu}); \check{\Gamma}=\frac{-1}{4} \{ \gamma_a, S_{(b)(c)}  \} e^{(a)\mu}e^{(b)\nu}\partial_{\mu}e^{(c)}_{\nu}.
\ee 
\be
\gamma^{(0)}=\begin{pmatrix}
0 & i \\
i & 0
 \end{pmatrix}\,,~~
 \gamma^{(i)}=\begin{pmatrix}
0 & i\sigma^i \\
-i\sigma^i & 0
 \end{pmatrix}\,,~~
\gamma_5=\begin{pmatrix}
1 & 0 \\
0 & -1
 \end{pmatrix}
 \ee 
Further, we define $\Psi_{R, L}=\frac{1}{2}(1\pm \gamma_5)\Psi$
and adopt $\Psi_{R,L} = \psi_{R,L} e^{-i\omega t+i\ell\varphi+ikz}$.
We work in a probe limit, i.e., in which the fermionic fields decouple from gravity.
Then, we have the following equations for $\Psi_1$ :
\begin{align}
\Big( \omega\gamma^{(0)}  +i\gamma^{\rho} \partial_{\rho} +\gamma^{\varphi} (\ell -\gamma_B k -\nu    ) -\gamma^{(3)} k   \Big)  \Psi_{L}^{(1)} &= (-ir^2\gamma_5 \partial_r +mr) \Psi_{R}^{(1)}, \nonumber \\
\Big( \omega\gamma^{(0)}  +i\gamma^{\rho} \partial_{\rho}+\gamma^{\varphi }(\ell -\gamma_B k  -\nu  ) -\gamma^{(3)} k  \Big)  \Psi_{R}^{(1)} &= (-ir^2\gamma_5 \partial_r +mr) \Psi_{L}^{(1)} ,
\end{align} 
where $\gamma^{\rho}=\gamma^{(1)}\cos \varphi + \gamma^{(2)} \sin \varphi $, $\gamma^{\varphi}= -\gamma^{(1) } \sin \varphi +\gamma^{(2)}\cos \varphi$, $A_{\varphi}=\Phi /2\pi$ and $\nu = \Phi /(\text{flux quantum})$. 
%%%%%%%%%%%%
%%%%%%%%%%%% 
%%%%%%%%%%%%
%%%%%%%%%%%%
We expand the bulk Dirac field $\Psi_1$ as follows:
\be
\Psi_R^{(1)}=
\begin{pmatrix}
U(r)_+u(\rho)_+ \\
V(r)_+v(\rho)_+ \\
S(r)_+s(\rho)_+ \\
W(r)_+w(\rho)_+ 
 \end{pmatrix},
 \Psi_L^{(1)}=
\begin{pmatrix}
U(r)_-u(\rho)_- \\
V(r)_-v(\rho)_- \\
S(r)_-s(\rho)_- \\
W(r)_-w(\rho)_- 
 \end{pmatrix},
\ee
then we have the following:
\begin{align}\label{D_Dirac}
\{\partial_{\rho}^2+\frac{1}{ \rho}\partial_{\rho} -\frac{1}{\rho^2}(\ell -\gamma_B k -\nu )^2+\omega^2 -k^2 \} u_{\pm},v_{\pm},s_{\pm},w_{\pm} =const, \nonumber \\
\Big( -r^4\partial_r^2-2r^3\partial_r +m^2r^2 \Big) U_{\pm} +\Big( \pm 2mr^3 \partial_r \pm mr^2  \Big)V_{\pm} =const,\nonumber \\ 
\Big( -r^4\partial_r^2-2r^3\partial_r +m^2r^2 \Big) V_{\pm} +\Big( \mp 2mr^3 \partial_r \mp mr^2  \Big)U_{\pm} =const,\nonumber \\  
\Big( -r^4\partial_r^2-2r^3\partial_r +m^2r^2 \Big) S_{\pm} +\Big( \mp2mr^3 \partial_r \mp mr^2  \Big)W_{\pm} =const,
\nonumber \\   
\Big( -r^4\partial_r^2-2r^3\partial_r +m^2r^2 \Big) W_{\pm} +\Big( \pm 2mr^3 \partial_r \pm mr^2  \Big)S_{\pm} =const.\end{align}
%%%%%%%%%%%%%%%%%%%%%
%%%%%%%%%%%%%%%%%%%%%
%%%%%%%%%%%%%%%%%%%%%  
We have the following equations for zero frequency from the Eq.(\ref{D_Dirac}).
\begin{align}\label{rho-part}
&(\Delta_{\ell- \nu- \gamma_B k}-k^2)u_{\pm}, v_{\pm},s_{\pm}, w_{\pm}  =0. 
\end{align}
where $\Delta_{t}=\partial_{\rho}^2+\frac{1}{\rho}\partial_{\rho}-\frac{t^2}{\rho^2}$. 
By choosing $\lvert \ell -\nu -\gamma_Bk \rvert=\frac{1}{2}$, i.e., $\ell =0$ and 
$\nu + \gamma_Bk=1/2$,     
these equations are similar to the solutions in the weak coupling system \cite{IT2019}.
\begin{equation}\label{zm01-2}
u_{\pm}, v_{\pm}, s_{\pm}, w_{\pm} \sim
\frac{e^{-\frac{\lvert \nu-1/2\rvert \rho}{\gamma_B} } e^{\frac{-i(\nu-1/2)z}{\gamma_B}}  }{\sqrt{\rho}}
\begin{pmatrix} 
i, &-e^{i\varphi}, & i, & -e^{i\varphi}
\end{pmatrix}^T.
\end{equation} 
%%%%%%%%%%%%%%%%%%%
For $\Psi_2$, by changing $\nu \rightarrow -\nu$, $k \rightarrow -k$ and by choosing
$\lvert  \ell +\nu +\gamma_B k \rvert=1/2$, i.e., $\ell=0$ and $\nu +\gamma_B k=1/2$, we also have the same solutions as Eq. (\ref{zm01-2}).
%%%%%%%%%%%%%%%%%%%
%%%%%%%%%%%%%%%%%%%
From the above solutions, the bulk fermion's zero frequency solutions are hosted by  $\rho =0$.
However, if we adjust the magnetic flux to change the value $\lvert  \ell- \nu- \gamma_B k    \rvert  >1 $, 
zero-modes are not hosted by $\rho =0$. The reason is given as follows: 
Near $\rho =0$, Eq.(\ref{rho-part}) has asymptotic solutions $\rho^{\lvert \ell -\nu -\gamma_Bk  \rvert}$ and 
$\rho^{-\lvert \ell -\nu -\gamma_Bk  \rvert}$. 
Because we consider finite norm based upon finite energy, we require square-integrable near $\rho =0$ \cite{Ishibashi-Hosoya1999}. Taking this consideration to the above solutions, $\rho^{-\lvert \ell -\nu -\gamma_Bk  \rvert}$ is excluded for $\lvert \ell -\nu -\gamma_Bk  \rvert >1.$  
From this fact, zero frequency solutions are not hosted by $\rho =0$. 
$\rho =0$  is the dislocation line in the boundary.    
The dislocation line maps into the gravitational bulk as 
a domain wall along the AdS radial direction, where the domain wall
hosts the zero frequency solution shown in Eq.(\ref{zm01-2}).

Then it is possible to set the five-dimensional theory \cite{Landsteiner-Liu2015,Tong_GT,Hill2006, Hughes et al 2016} anew by action $S=S_0+S_1+S_{CS}+S_{DW}+S_{\text{boundary}}$, where  
\begin{align}
S_{CS}&=
c_1 \int d^5x  \varepsilon^{\alpha\beta\eta\mu\sigma}A_{\alpha}\partial_{\beta}A_{\eta}\partial_{\mu}A_{\sigma}, 
\nonumber \\
S_{DW}&=c_2\int_{DW}d^4x \bar{\psi}_1(iD-m) \psi_1+\bar{\psi}_2(iD+m) \psi_2 .  
\end{align}
In the above expression,  
$S_{CS}$ is the topological Chern-Simons term, and 
$S_{DW}$ is the fermionic matter action on the domain wall, where $D$ is the covariant derivative on the domain wall, and $\psi_1$ and $\psi_2$ 
correspond to $\Psi_1$ and $\Psi_2$, respectively. 
In the following, we consider the gauge field as the background gauge field. 
If we consider a gauge transformation: 
\be
A_{\alpha} \rightarrow A_{\alpha}+\partial_{\alpha}\theta, 
\ee  
$S_{1}$ is strictly invariant. 
On the other hand, the following relation can be derived for $S_{CS}$:
\be\label{gt_CS}
S_{CS}\rightarrow 
S_{CS}-c_1\int_{}d^5x \partial_{\alpha} \Big(  \varepsilon^{\alpha\beta\eta\mu\sigma} 
\theta \partial_{\beta}A_{\eta}\partial_{\mu}A_{\sigma}  \Big).
\ee
We cannot discard the total derivative term in the domain wall's presence, 
which hosts the zero frequency solution at $\rho =0$.
Further, for $S_{DW}$, 
\be
S_{DW}\rightarrow
S_{DW}+c'_2\int_{\rho =0}d^4x \bar{\psi}_2\gamma_{\mu}\partial^{\mu}\theta \psi_2.
\ee 
This is because that $\psi_1$ and $\psi_2$ are opposite in mass and axial charge. 
Only when the $c_1$ and $c'_2$ satisfy a specified relationship, 
bulk becomes gauge-invariant. 
This correspondence is similar to anomaly cancellation in compactified extra dimensions \cite{Hill2006, Gripaios-West2008}. 
From this point of view, 
$\rho =0$ in the boundary is reproduced in bulk 
as a domain wall along the AdS radial direction, making the theory 
not gauge invariant.
Then, we need the Chern-Simons term to maintain the gauge invariance.

However, Eq.(\ref{gt_CS}) gives the anomalous current to the boundary, for
the bulk theory should be invariant for gauge transformations that do not vanish at infinity, where    
the domain wall exists. Since the fermion anomaly cancels this anomalous current, 
the holographic principle realizes anomaly inflow. 
In summary, Chern-Simons terms in the bulk describe 't Hooft anomalies \cite{Benini2018}.

Even if we leave the violation of gauge invariance, the GKP-Witten relation
\be\label{GKP-Witten01}
\langle exp(i \int_{\text{boundary}} A_{\mu}J^{\mu}  )  \rangle = e^{iS[A_{\mu}]_{\text{on-shell}}    } 
\ee 
gives the $\partial_{\mu} \langle  J^{\mu}  \rangle \ne 0   $.
The bulk's gauge symmetry breaking generates the non-conserved current to the boundary.  
This situation also leads to the generation of the 't Hooft anomaly.

In both cases, the generation of the 't Hooft anomaly on the boundary is realized by the holographic model.   
From Eq. (\ref{zm01-2}) and its explanation, by changing the value from $\lvert \ell -\nu -\gamma_Bk \rvert <1$ to 
$\lvert \ell -\nu -\gamma_Bk \rvert  > 1$, zero frequency solution in no longer hosted by  $\rho = 0$ and vice versa. This fact shows that the bulk's gauge field adjusts the domain wall creation/annihilation.
As a result, we can control the generation of the 't Hooft anomaly.

't Hooft anomaly on the boundary indicates 
 three aspects. 
The first aspect is the generation of the singularity of the boundary.
We consider Euclidean time evolution of the boundary $B$.  
The path integral on  
is given by Euclidean time evolution  $e^{-\varepsilon H}$,
where $H$ is the Hamiltonian, and $\varepsilon$ is imaginary time. 
The ground state of the boundary $\lvert  \Omega \rangle$ changes adiabatically.   
The time-constant surface can be recognized as the border of the space and corresponds to 
the state $\lvert  \mathfrak{B}  \rangle$.
Consequently, the partition function of the boundary is given by 
\be
Z=\langle  \mathfrak{B} \vert \Omega \rangle.
\ee
The 't Hooft anomaly emerges from the phase ambiguity of the partition function \cite{Witten-Yonekura2019}.
The singularity in the parameter space generates the phase ambiguity of the partition function for the dislocated boundary, 
preventing smooth gauge fixing.
This singularity is specified as a set of the vanishing point of the ground state $\lvert \Omega \rangle$ in parameter space: 
i.e., 
This set is a curved line defined by an intersection of two curved surfaces  $Re \lvert \Omega \rangle = Im \lvert \Omega \rangle =0 $ in parameter space. 
The connection $A=\langle \Omega \vert d\Omega \rangle$ is defined, and
the intersection of $Re\lvert \Omega \rangle =Im\lvert \Omega \rangle=0 $ allows the wave function's phase difference is equal to an 
integral multiple of $2\pi$ for the closed curve $\mathcal{C}$ around it:
$\oint_{\mathcal{C}} A d\lambda + 2\pi n$. 
When the closed curve is shrunk to the point,
$\oint_{\mathcal{C}} A d\lambda = \int \mathbf{B}\cdot d\mathbf{S}    = - 2\pi n$.
This situation is similar to the singular line, which endpoint is the point magnetic charge.  
Namely, the intersection corresponds to the Dirac string 
\cite{Hatsugai2010}. 
As a result, the boundary is characterized by a topological invariant.
Whereas pure AdS case, there is no phase ambiguity. This is because the corresponding holographic model 
will not break the gauge invariance on the bulk spacetime \cite{Liu-McGreevy-Vegh2011}.   
%%%%%%%%%
%%%%%%%%%

From the fact that 't Hooft anomaly is renormalization group invariant,  
the existence of 't Hooft anomaly in UV theory assures  
that in IR effective theory. 
When the symmetry acts gapped nondegenerated ground state trivially, an anomaly is not generated. 
As a result, this type of ground state is not approved by the IR theory and has a degeneracy originated from spontaneous symmetry breaking or 
topological order. 
In this case, all the Berry curvature emanated from the UV layer flows to the IR layer, and 
the total Berry flux is conserved in each layer \cite{WCLGQR2016}. 
In other words, there is no topological phase transition in the direction of entanglement renormalization. 
The topologically nontrivial state of the bulk corresponds to the domain wall generation of the bulk spacetime.

The second aspect is that the anomaly inflow is related to the phase transition. 
Bulk gauge fields control the domain wall creation and annihilation
to cause the phase transition. In response to this change, the Chern-Simons term contributes to the gauge invariance of the bulk. At the same time, the Chern-Simons term generates a current $J_{\mu} = \frac{\delta S_{CS} [A] }{\delta A_{\mu}}$ on the four-dimensional boundary. 
%
%The current has a non-vanishing divergence on the four-dimensional boundary where it is cancelled by the anomaly. This mechanism is referred to as anomaly inflow. 
%
The current has a non-vanishing divergence on the four-dimensional boundary where the anomaly cancels it.  
Conversely, this anomaly inflow 
realizes the phase transition.

%%%%%%%%%
%%%%%%%%% 

The third aspect is the entropy production on the boundary.
The generation of 't Hooft anomaly shows 
the phase transition and can be related to entropy production.
Because the anomaly generation parallels the domain wall's creation, the entropy production occurs at $\rho =0$ where the domain wall exists.  
This entropy can be calculated from the holographical methods:
In the AdS/CFT correspondence, the entanglement entropy 
for a region $A$ is obtained from the minimal surface $\gamma_A$ 
in bulk geometry ends at $\partial A$ \cite{Ryu-Tnayanagi2006}. 
\be
S_A = \frac{Area(\gamma_A)}{G_N},
\ee
where $G_N$ is the bulk Newton constant. This procedure applies to the presence of a defect or boundary \cite{Kobayashi-Nishioka-Sato-Watanabe2019}. 
However, we have to take into consideration that the holographic configuration is not changed
regardless of the existence of the domain wall. 
From this viewpoint, the change of the boundary condition at $\rho =0$ causes entropy production.  

$\rho =0$ is an entangling surface defined as a boundary between the regions.
For any QFT, Dirichlet or Neumann boundary condition is selected for a weakly coupled system. However, in a strong coupling system, we do not have the natural choice of the boundary condition.
The fact $Re\lvert \Omega \rangle =Im\lvert \Omega \rangle =0$ 
suggests that the dislocation line provides a specific boundary condition  
for the formation of the singular line in parameter space.  
Further, this boundary condition leads to the generation of topological invariants. 
Expressly, this boundary condition relates to the topological invariant around the 
singularity in the parameter space.

This leads to the situation that entropy production can be generated from the non-universal form \cite{Ohmori-Tachikawa2015}. 
This non-universal form characterizes the quantum singularity in the strong coupling system. 
For detailed consideration, this is a future consideration.  

Another thing to consider is a quantum correction to the holographic entropy. 
In this correction, bulk entanglement entropy contributes to the von Neumann entropy and 
corresponds to the generalized entropy in black hole thermodynamics \cite{FLM2013}.
The minimal surface corresponds to the boundary's dislocation line and spreads on the domain wall.
The existence of the domain wall shows that the UV state's nontrivial topological state 
leads to the state in each layer at the bulk is similarly nontrivial.  

Because the domain wall provides a topologically nontrivial state in bulk, it generates more long-range entanglement entropy than without it. 
This entropy constitutes the bulk entanglement entropy which contributes to 
holographic entanglement entropy. This effect gives the clue for understanding by the 
lattice MERA.

To generate the long-range correlation in lattice MERA, 
we must prepare a larger number of the layer. 
From this fact and the physics is limited by 't Hooft anomaly, the domain wall has an effect in the renormalization direction. 
Fig. 2 shows the MERA network. 
For detailed consideration, this is also a future consideration.   
%%%%++++++++
%To generate the long range correlation in lattice MERA, larger number of the layer is needed. 
%From this effect in accordance with the physics is limited by 't Hooft anomaly,  
%the domain wall weakens the effect of coarse-graining. 
%Fig. 2 shows the MERA network. 
%For detailed consideration, this is also a future consideration.  
%%%%+++++++
%in the bulk and generates  long-range entanglement entropy, it weakens the effect of disentanglement and coarse-graining. 

\begin{figure}[htbp]
  \begin{center}\label{MERA}
   \includegraphics[width=100mm]{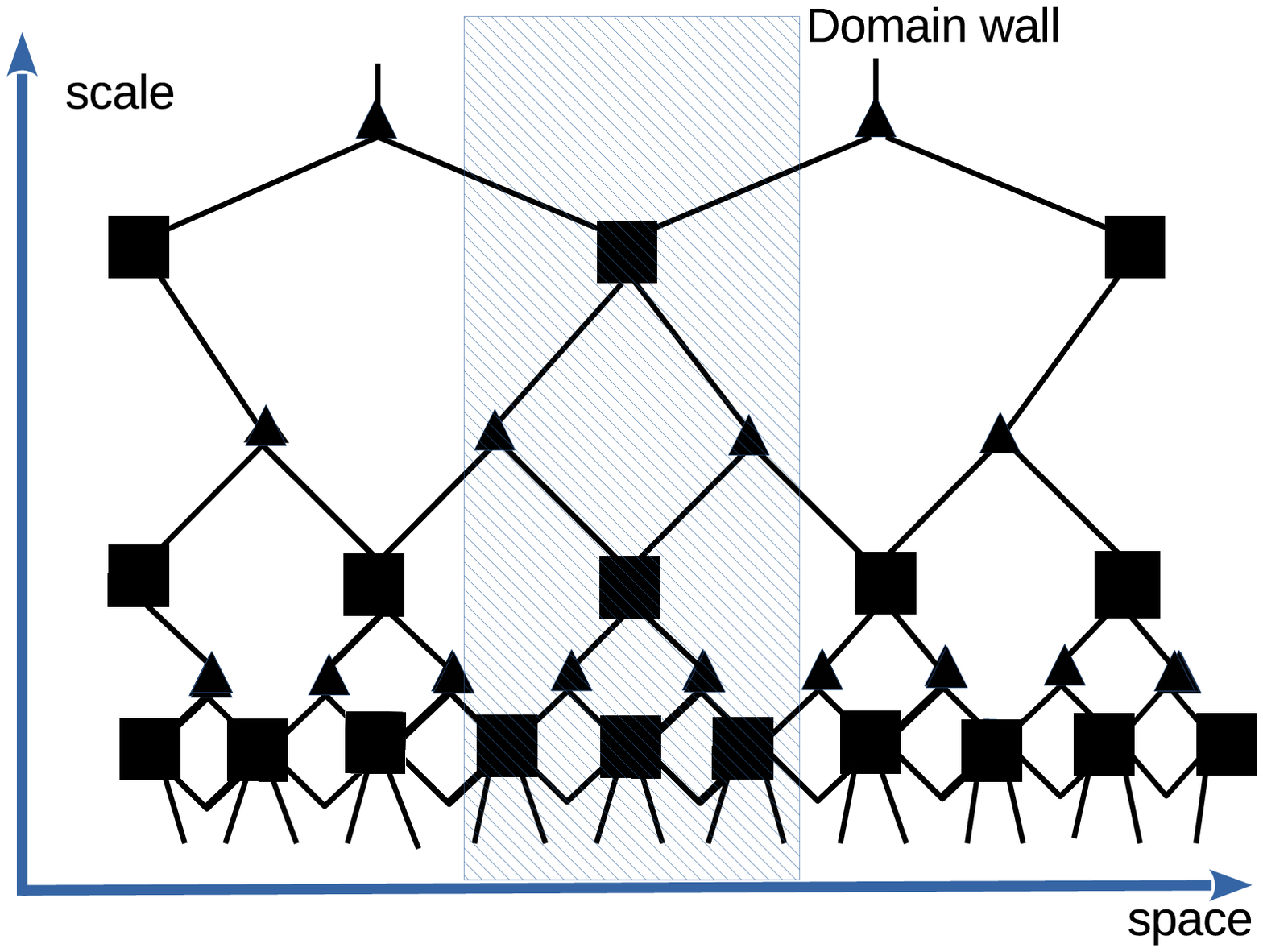}
  \end{center}
  \caption{The MERA network under the influence of the domain wall}
%\footnotemark[1]
  \label{fig:PhaseT}
\end{figure}

Our situation shares the case with global inconsistency \cite{GKKS2017}: 
Global inconsistency exists when changing the parameter of the system and gauging discrete symmetry and global symmetry, i.e., there is a discontinuous change in the fundamental property of the system in this process. A 't Hooft anomaly is the obstruction for gauging symmetries, and global inconsistency is a milder obstruction than the 't Hooft anomaly. Global inconsistency for a quantum mechanical system involving Chern Simons terms has been analyzed.
There is a case that the system needs the Chern Simons term with the different value of coefficients in the region of continuous parameter space to conserve certain symmetry for global inconsistency \cite{Kikuchi-Tanizaki2017}. In our situation, we need Chern Simons term to preserve gauge transformation, which relates to the generation of 't Hooft anomaly. 
$\nu$ and $\gamma_B$ is the parameter of the theory. 
Moreover, to maintain the bulk's gauge invariance, we need to adjust the value of the coefficient $c_1$.

%%%%%%%%%%%%%%%%%%%%%%%%%%%
%%%%%%%%%%%%%%%%%%%%%%%%%%% 
%%%%%%%%%%%%%%%%%%%%%%%%%%
\section{Conclusion and discussion}
\label{}
%%%%%%%%%%%%%%%%%%%%%%%%%%%
%%%%%%%%%%%%%%%%%%%%%%%%%%%

In \S 2, the dislocation generates the quantum singularity, related to the breaking of topological material's translational symmetry. % \textbf{ for the topological. }
For the topological material, the quantum singularity creates the defect in the momentum space ubiquitously and leads to the phase transition. 
Moreover, the kinetic equation reveals that the defect generates an anomaly without the characteristic energy scale.

In \S 3, the quantum singularity is imported into the QFT through topological protection. 
Accordingly, the three-dimensional dislocations map into the gravitational bulk as domain walls extending along the AdS radial direction from the boundary in the holographic model. 
The domain wall in the bulk spacetime needs the Chern-Simons term for maintaining the gauge invariance. 
In response, 
't Hooft anomaly is generated on the boundary. 
The creation/annihilation of the domain wall causes the quantum phase transition by 't Hooft anomaly generation and is controlled by the gauge field.
In other words, the phase transition is realized by the anomaly inflow.

This 't Hooft anomaly is caused by a phase ambiguity of the ground state. 
This ambiguity comes from the singularity, specified as a set of the ground state's vanishing point in parameter space. 
The singularity prevents the smooth gauge fixing and 
gives the basis for the boundary's topological state with the Berry connection.  
't Hooft anomaly's renormalization group invariance shows that   
all the Berry curvature emanated from the UV layer flows to the IR layer, and the total Berry flux is conserved in each layer. The topologically nontrivial state of the bulk corresponds to the domain wall generation of the bulk spacetime.

In response to domain wall creation on the bulk, the entropy is produced around the boundary $\rho =0$. 
The entropy can be specified by the boundary condition at $\rho =0$ and produced from the non-universal form. 
This non-universal form characterizes the quantum singularity in the strong coupling system. 
Another aspect is that the bulk entanglement entropy gives the quantum correction to the holographic entropy. 
%This effect also can be understood from lattice MERA.
In summary, the holographic quantum singularity is characterized by the parameter space's singularity and the nature of the entropy.

\appendix\label{apdx}

%%%%%%%%%%%%%%%%%%%%%%%%%%%%%%%
%%%%%%%%%%%%%%%%%%%%%%%%%%%%%%%
\section{Quantum singularity in topological insulators}
%%%%%%%%%%%%%%%%%%%%%%%%%%%%%%%
%%%%%%%%%%%%%%%%%%%%%%%%%%%%%%%
%%%%%%%%%%%%%%%%%%%%%%%%%%%%%%%
%%%%%%%%%%%%%%%%%%%%%%%%%%%%%%%
%%%%%%%%%%%%%%%%%%%%%%%%%%%%%%%
%\subsection{Topological state on dislocated media}
%%%%%%%%%%%%%%%%%%%%%%%%%%%%%%%
%%%%%%%%%%%%%%%%%%%%%%%%%%%%%%%
%%%%%%%%%%%%%%%%%%%%%%%%%%%%%%% 

In the following, we summarize  
the result already obtained \cite{IT2019}. 
We consider the following dislocated metric, which describes the static dislocated spacetime
\cite{Galtsov-Letelier,Furtado-Berezza-Moraes2001}:
\begin{equation}\label{GLT01}
ds^2=-dt^2+dr^2+\rho^2d\varphi +(\gamma_B d\varphi +dz)^2, 
\end{equation}
where $2\pi\gamma_B$ is analogous to the Burgers vector.  
We consider a modified Dirac equation describing the topological state \cite{Shen2012}.  
\begin{align}
&\Big( i \gamma^{(0)} \partial_t + i \gamma^{(\rho)} \partial_{\rho} +\frac{i}{\rho} \gamma^{(\varphi)} (\partial_{\varphi} -\gamma_B \partial_z ) +  \nonumber \\
&i \gamma^{(3)} \partial_{z} -\{M_0+M_2(p_x^2+p_y^2+p_z^2) \} \Big) \Psi = 0.  
\end{align}  
Following representation of the Dirac matrices is used:
\begin{align} 
&\gamma^{(0)}= 
\left(    \begin{array}{cc}
-i\sigma^3 & 0  \\ 
0 & i\sigma^3
\end{array} \right), \hspace{1mm}
\gamma^{(1)}= 
\left(    \begin{array}{cc}
\sigma^2 & 0  \\ 
0 & -\sigma^2
\end{array} \right) \nonumber \\
&\gamma^{(2)}= 
\left(    \begin{array}{cc}
-\sigma^1 & 0  \\ 
0 & \sigma^1
\end{array} \right), \hspace{1mm}
\gamma^{(3)}= 
\left(    \begin{array}{cc}
0 & -i  \\ 
i & 0
\end{array} \right).
\end{align}             
In these expressions, $\sigma^1$, $\sigma^2$, and $\sigma^3$ are the usual Pauli 
matrices.

In the above equation,  we need to express $p_x^2+p_y^2+p_z^2$ in the dislocated metric.
States $M_2=0$ corresponds to a trivial topological insulator and further when $M_0=0$ corresponds to gapless semimetal. 
The phase diagram without the dislocation has already known and shown as Fig. {\ref{fig:PhaseT}}(A). 
The phase diagram features the vacuum of Standard Model 
\cite{Volovik_SM}.

Following anzats as a wave function for 
the modified Dirac system 
\begin{equation}
\Psi =
\begin{pmatrix}
\sqrt{E+M_0}v_1(\rho)\\
\sqrt{E+M_0}v_2(\rho)ie^{i\varphi}
\end{pmatrix}
exp(-iEt+i\ell\varphi +ikz),
\end{equation}
%%%%%  
the equation for radial direction is given as follows: 
\begin{equation}\label{radial03}
v_j''+\frac{1}{\rho}v_j'+\Big( -\kappa-\frac{(\tau+j-1)^2}{\rho^2}    \Big)v_j=0;  \hspace{1mm} j=1,2 ,
\end{equation} 
where $\tau=\ell-\gamma_Bk$. 
$\kappa $ is given by the following: 
\begin{align}
\kappa &= \frac{(2M_2M(k)+1)\pm\sqrt{1+4(M_2E)^2+4M_2M(k)}}{2M_2^2}, 
\end{align}
where $M(k)=M_0+k+M_2k^2$. 
Then, the solutions are given as Bessel and Neumann functions:
\begin{equation}\label{TC01}
v_j(r) = C_1 J_{\lvert \tau+j-1 \rvert}(-\sqrt{-\kappa}\rho)+C_2 N_{\lvert \tau+j-1 \rvert} (-\sqrt{-\kappa}\rho), \hspace{1mm} j=1,2.
\end{equation}
%%%%
The quantum singularity creates the region $|\tau +j-1|< 1, \hspace{1mm} (j=1, 2)$ as a kind of defect in the momentum space.  
The reason for this is that the region $|\tau +j-1|< 1$ corresponds to the not essentially self adjoint operator for the above equation, and 
the region $|\tau +j-1|\ge 1$ corresponds to the essentially self-adjoint operator.
Then, two regions correspond to different physics.

Because the deficiency indices are given as $(1, 1)$, self-adjoint extensions should be considered.
A four-parameter family of self-adjoint extension is as follows: 
\begin{equation}\label{SAE03}
\begin{pmatrix} 
\phi_1 \\ \phi_3
\end{pmatrix}
=
\begin{pmatrix} 
p & q+ir \\ -q+ir & s
\end{pmatrix}
\begin{pmatrix} 
\phi_2 \\ \phi_4
\end{pmatrix}
,\hspace{2mm}\rho \rightarrow 0
\end{equation} 
and 
\begin{equation}\label{SAE04}
\begin{pmatrix} 
\psi_1 \\ \psi_3
\end{pmatrix}
=
\begin{pmatrix} 
p & q+ir \\ -q+ir & s
\end{pmatrix}
\begin{pmatrix} 
\psi_2 \\ \psi_4 .
\end{pmatrix}
,\hspace{2mm}\rho \rightarrow 0,
\end{equation} 
where $p$, $q$, $r$, and $s$ are real numbers. 

For $ M_2M(k)  >-1/4$;     
if $\lvert \tau +j-1  \rvert =\frac{1}{2}$ and $\ell =0$, $\gamma_Bk=\frac{1}{2}$ is obtained.
From the overcomplete basis, 
self-adjoint extensions of linear combinations are considered.  
The linear combinations are formed by states proportional to $e^{-\kappa_+ \rho}$ or $e^{-\kappa_- \rho}$. 
A zero-energy state bound to the dislocation is obtained with $p=q=r=s=1$: 
\begin{equation}\label{solDS001}
\Phi \sim \frac{(e^{-\kappa_+ \rho}-e^{-\kappa_- \rho})e^{-iz/\gamma_B}}{\sqrt{\rho}}\left( \begin{array}{cccc}
i, & -e^{i\varphi},  &  i ,   &  -e^{i\varphi}\end{array} \right)^T, \hspace{1mm} \kappa_{\pm}=\frac{ 1 \pm \sqrt{ 1+ 4M_2M(k)   }  }{2\lvert M_2 \rvert}
\end{equation}

For $ M_2M(k)  < -1/4$; if $\lvert \tau +j-1  \rvert =\frac{1}{2}$ and $\ell =0$, $\gamma_Bk=\frac{1}{2}$ is obtained. 
From the overcomplete basis, self-adjoint extensions of linear combinations
are considered. The linear combinations are formed by the states proportional to $e^{-\frac{1}{2B} \rho}$ .
A zero-energy state bound to the dislocation is obtained with $p=q=r=s=1$:  
\begin{equation}\label{solDS002}
\Phi \sim \frac{e^{-\alpha\rho}e^{-iz/\gamma_B}\sin \beta \rho}{\sqrt{\rho}}\left(    \begin{array}{cccc}
i, & -e^{i\varphi},  &  i ,   &  -e^{i\varphi}\end{array} \right)^T.
\end{equation}
where $\alpha=1/2\lvert  M_2 \rvert$ and $\beta=\sqrt{-1-4M_2M(k)}/2\lvert  M_2 \rvert$. 

For $M_2M(k)  =-1/4$, there is no bound solution which is regular at $\rho =0$.

Next, a topological invariant originated in the dislocation is considered.  
The existence of quantum singularity requires
full quantum treatment of the topological number characterizing the
topological invariant.
The topological invariant is calculated by the integration of single particle Green's function over momentum space in inhomogeneous systems.  
\begin{align}\label{TN02}
&N=\frac{1}{24\pi}
\int_{-\infty}^{\infty} 
d\mathbf{p}^3 \int_{-\infty}^{\infty} 
d\omega   \nonumber \\
&\varepsilon^{\mu\nu\rho\sigma}
tr\Big(  G^{-1}\partial_{\mu}G G^{-1}\partial_{\nu}G G^{-1}\partial_{\rho}G  G^{-1}\partial_{\sigma}G
   \Big),   
\end{align}
where $(\mu, \nu, \rho, \sigma)$ run over $(\omega, p_x, p_y, p_z)$  
and $\varepsilon$ is a fully antisymmetric tensor.   
Single particle Green's function $G$ is defined by Dyson's equation  \cite{Shiozaki-Fujimoto2012}.

Dyson's equation defines single-particle Green's function $\tilde{G}_{\alpha\beta}(i\omega, \mathbf{x}_1, \mathbf{x}_2)$: 
\begin{equation}
\int d\mathbf{x}_2 \tilde{K}(i\omega, \mathbf{x}_1, \mathbf{x}_2)\tilde{G}(i\omega, \mathbf{x}_1, \mathbf{x}_2)=\delta(\mathbf{x}_1-\mathbf{x}_2),
\end{equation}
where $\tilde{K}(i\omega, \mathbf{x}_1, \mathbf{x}_2)=[i\omega -H(-i\nabla_1, \mathbf{x}_1)]\delta (\mathbf{x}_1,\mathbf{x}_2)$ is the kernel of Green's function.
Then   
$\tilde{G}(i\omega, \mathbf{x}_1, \mathbf{x}_2)$ gives
\begin{equation}
G(i\omega, \mathbf{p})=
\int d\mathbf{r} \tilde{G}(i\omega, \mathbf{x}_1, \mathbf{x}_2)e^{-i\mathbf{p}\cdot \mathbf{r}}. 
\end{equation}
In the above expression, $\mathbf{p}$ is the momentum of the relative coordinate $\mathbf{r}=\mathbf{x}_1-\mathbf{x}_2$. 
For the translational symmetric case, $G$ is the matrix inverse of $K$, $G=K^{-1}$, but in the case of inhomogeneous systems, $G\ne K^{-1}$.

Low-energy approximation requires its momentum $\mathbf{p}$ to lie in shell width $\Delta p$ around the specific value, and
its frequency is smaller than cut-off energy $\omega_c$. 
Values of $\Delta p$ and $\omega_c$ do not contribute to physical results.      
From this viewpoint, the domain of integration, i.e., the base space, can be set out as following Fig.\ref{fig:one}. 
This space is a suspension of a cylinder and homotopic to $S^3$.

Accordingly, the topological number characterizes the homotopy of the map 
$G: (\omega, p_x, p_y, p_z) \mapsto G(\omega, p_x, p_y, p_z)\in GL(n, \mathbb{C})$, where $n$ is the number of the band, i.e., $G(\omega, p_x, p_y, p_z) $ defines a map from the $S^3$ momentum space to  the space of non-singular Green's function. This space belongs to the 
group $GL(n, \mathbb{C})$, whose homotopy group is labelled by an integer: 
$\pi_3(GL(n,\mathbb{C}))=\mathbb{Z}$.

If there is no dislocation, i.e., $\gamma_B=0$, the following relation is known from the semiclassical calculation where 
$G(i\omega, \mathbf{p} )=[i\omega-H(\mathbf{p})]^{-1}$ :
\begin{align}\label{tnumber01}
    N
=  \text{sgn}(M_0)-  \text{sgn}(M_2) 
\end{align}
Fig. \ref{fig:PhaseT}(A) describes this relation.
For example, if $M_0M_2<0$, then $N=2$, i.e., non-trivial state \cite{Shen2012}. 
Further, if $M_0=M_2=0$, massless Dirac fermion is obtained, where equal numbers of right and left fermionic species.
%%%%
\begin{figure}[htbp]
 \begin{minipage}{0.4\hsize}
  \begin{center}
   \includegraphics[width=50mm]{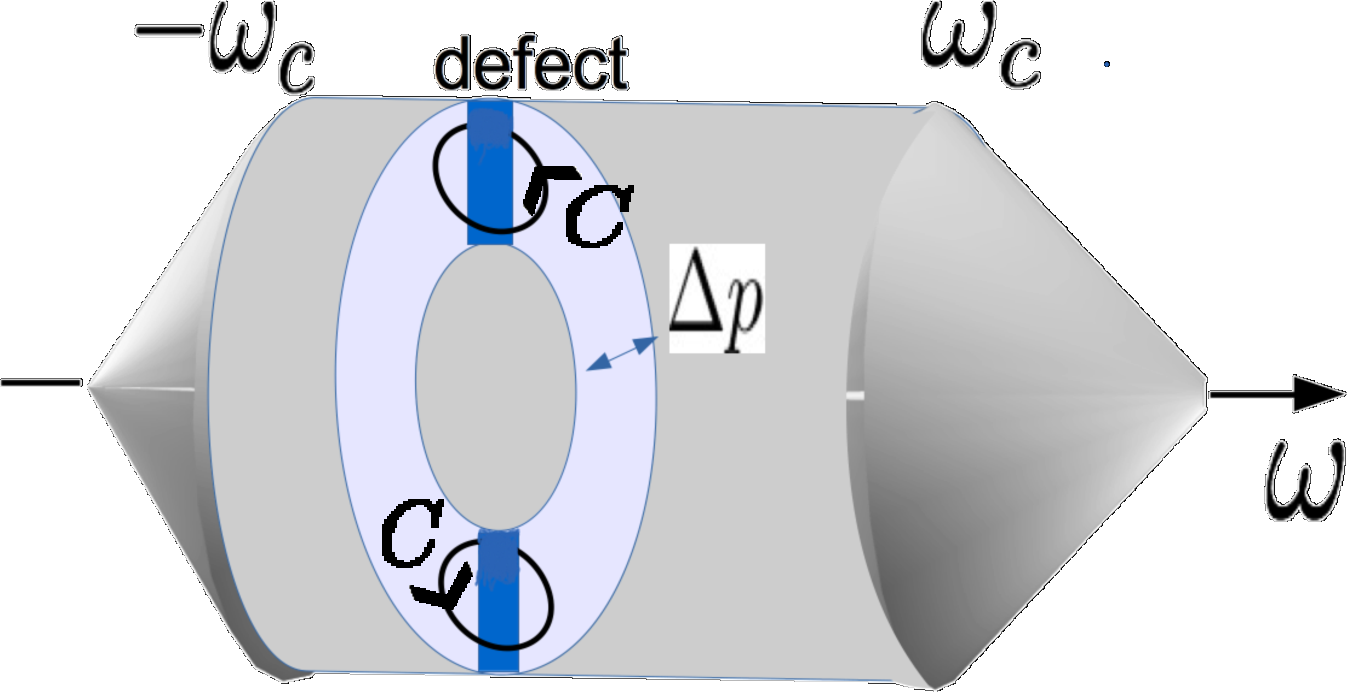}
  \end{center}
  \caption{Base space in low-energy approximation}
  \label{fig:one}
 \end{minipage}
 \begin{minipage}{0.4\hsize}
  \begin{center}
   \includegraphics[width=60mm]{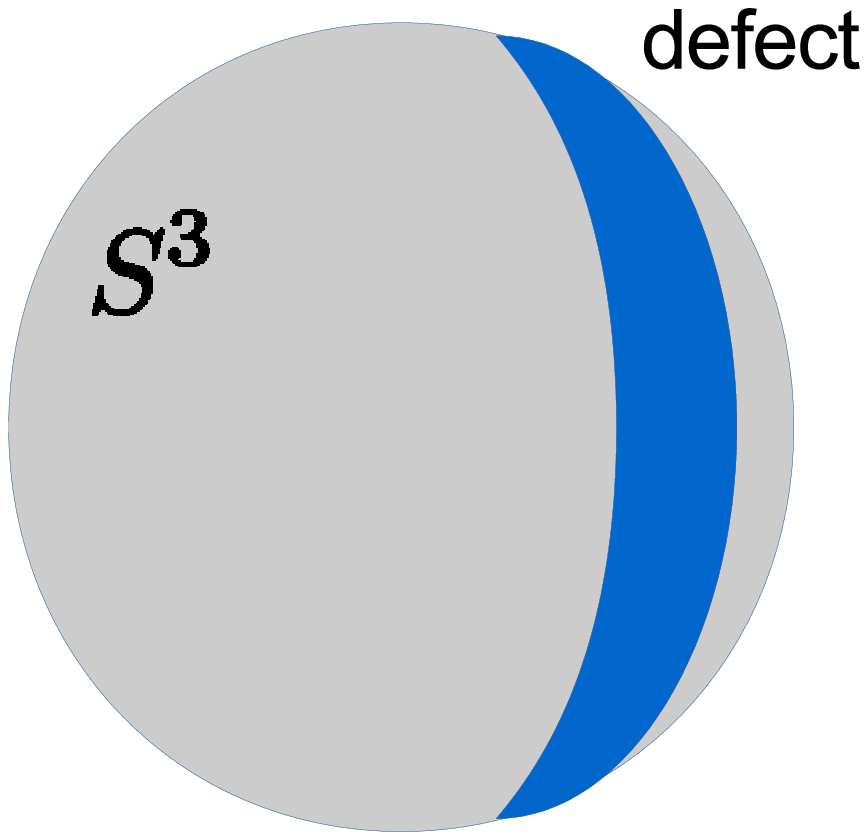}
  \end{center}
  \caption{Base space as a suspension of a cylinder}
  \label{fig:two}
 \end{minipage}
\end{figure}
%%%%%%%%%%
When calculating the topological invariant defined by
Eq.(\ref{TN02}),
the quantum singularity creates the region $\lvert \tau+j-1 \rvert <1$ as a kind of defect in the base space. 
The reason for this is that
the region $\lvert \tau+j-1 \rvert <1$ and 
the region $\lvert \tau+j-1 \rvert \ge 1$
correspond to different physics. 
From this defect, base space does not have the $S^3$ configuration, and the topological winding number 
cannot be defined, as shown in Fig. \ref{fig:two}. 

At the same time, the medium acquires a defect winding number given by 
\begin{equation}
\tilde{N}= \oint \frac{dl}{2\pi i}  G(i\omega, \mathbf{p})\partial_l G(i\omega, \mathbf{p}).  
\end{equation}  
The above integral is taken over an arbitrary contour $C$ enclosing the defect in the momentum space, as shown in Fig. \ref{fig:one}.   
The phase of the Green's function changes by $2\pi \hat{N}$ ($\hat{N}\in \mathbb{N}$) around the defect and edges of it are vortices with $\tilde{N}_1$ and $\tilde{N}_2$ where $\tilde{N}=\tilde{N}_1 +\tilde{N}_2$.  
$\tilde{N}_1$ and $\tilde{N}_2$ are circulation quantums of the edge and integers or half-integers. 
Because the zero-energy mode exists in the defect,
it is protected by these vortices and topologically stable. 
The regular zero-energy mode bound by the dislocation shows the existence of the quantum singularity.
From this viewpoint, all of the topological winding numbers change to zero, while simultaneously, a non-vanishing defect winding number is obtained. This situation is shown in Fig. {\ref{fig:PhaseT}}(B)

When an external magnetic field is applied to the system, 
Magnetic flux affects the condition of the quantum singularity. 
We consider the following vector potential
\be
A_{\varphi}=\frac{\Phi}{2\pi}.
\ee
This potential represents the magnetic vortex carrying the flux $\Phi$. 
By defining $\tau'=\ell-\nu-\gamma_Bk$ with $\nu = \frac{\Phi }{\text{magnetic flux quantum}} $, we can discuss a similar way 
as without the gauge field. 
If $\lvert \tau'+j-1 \rvert  \ge 1$ is not satisfied, applying the magnetic field makes $\lvert \tau'+j-1 \rvert  \ge 1$, and vice versa: 
By the existence of the quantum singularity, the defect is formed in the base space. Although this defect prohibits defining the winding number, it offers the topological number surrounding the defect. Whereas by adding the magnetic field, all modes can be removed from the region defined by $\lvert \tau'+j-1 \rvert  < 1$. 
Thus we can fine-tune the condition to control the quantum singularity by applying the magnetic flux. 
%%%%%%%%%%%%%%%%%%%
%%%%%%%%%%%%%%%%%%%

%%%%%%%%%%%%%%%%%%%%%%%%%%%%%%%%%%%%%%%%%%%%%%%%%%%
%%%%%%%%%%%%%%%%%%%%%%%%%%%%%%%%%%%%%%%%%%%%%%%%%%%
%%%%%%%%%%%%%%%%%%%%%%%%%%%%%%%%%%%%%%%%%%%%%%%%%%%
%%%%%%%%%%%%%%%%%%%%%%%%%%%%%%%%%%%%%%%%%%%%%%%%%%%


\begin{thebibliography}{00}%\label{sec:TeXbooks}


\bibitem{Volovik_SM}
G.~ E.~Volovik, Topological invariants for standard model: From semi metal to topological insulator,     {\it JETP Lett} \textbf{91} (2010),  55-61.
%%%%%%%%%%%%%%%%%%%%%%%%%%%%%%%%%%%%%%%%
%G.~E.~Volovik
%\emph{  Topological invariants for standard model: From semi metal to topological insulator, }
%JETP Lett \textbf{91} (2010) 55-61.
%https://doi.org/10.1134/S0021364010020013.
%%%%%%%%%%%%%%%%%%%%%%%%%%%%%%%%%%%%%%%%%%%%%%%%%%%% 
%%%%%%%%%%%%%%%%%%%%%%%%%%%%%%%%%%%%%%%%%%%%%%%%%%%% 
\bibitem{Maldacena}
J.~M.~Maldacena, The large N limit of superconformal field theories and supergravity, {\it Adv. Theor. Math. Phys.} \textbf{2} (1998), 231-252 ; The Large-N Limit of Superconformal Field Theories and Supergravity {\it Int. J. Theor. Phys. }\textbf{38} (1999), 1113-1133.

%%%%%%%%%%%%%%%%%%%%%%%%%%%%%%%%%%%%%%%%%%%%%%%%%%%
%%%%%%%%%%%%%%%%%%%%%%%%%%%%%%%%%%%%%%%%%%%%%%%%%%%
%%%%%%%%%%%%%%%%%%%%%%%%%%%%%%%%%%%%%%%%%%%%%%%%%%%
\bibitem{HHH2008}
S.~A. Hartnoll, C.~P. Herzog, and G.~T. Horowitz, 
Building a holographic superconductor,  
{\it Phys. Rev. Lett.}  \textbf{101} (2008), 031601. %%[0803.3295 [hep-th]].  

%@article{PhysRevLett.101.031601,  title = {Building a Holographic Superconductor},  author = {Hartnoll, Sean A. and Herzog, Christopher P. and Horowitz, Gary T.},  journal = {Phys. Rev. Lett.},  volume = {101},  issue = {3},  pages = {031601},  numpages = {4},  year = {2008},  month = {Jul},  publisher = {American Physical Society},  doi = {10.1103/PhysRevLett.101.031601},  url = {https://link.aps.org/doi/10.1103/PhysRevLett.101.031601}} 


%%%%%%%%%%%%%%%%%%%%%%%
%%%%%%%%%%%%%%%%%%%%%%%
%%%%%%%%%%%%%%%%%%%%%%%

\bibitem{Gubser-Pufu2008}
S. S. Gubser and S. S. Pufu,  
The gravity dual of a p-wave superconductor, 
%\emph{ The gravity dual of a p-wave superconductor, }
{\it JHEP }  {\bf 0811} (2008), 033. % [0805.2960 [hep-th]]. 
%
%S. S. Gubser and S. S. Pufu, \emph{ The gravity dual of a p-wave superconductor, }
%JHEP 0811 (2008) 033 [0805.2960 [hep-th]].
%https://doi.org/10.1088/1126-6708/2008/11/033.
%
%[16] K. Goldstein, N. Iizuka, S. Kachru, S. Prakash, S.P. Trivedi, and A. Westphal, J. High Energy Phys. 10 (2010)

%%%%%%%%%%%%%%%%%%%%%%%%%%%%%%%%%%%%%%%%%%%%%%%%%%%
%%%%%%%%%%%%%%%%%%%%%%%%%%%%%%%%%%%%%%%%%%%%%%%%%%%

\bibitem{Roberts-Hartnoll2008}
M. M. Roberts and S. A. Hartnoll, 
Pseudogap and time reversal breaking in a holographic superconductor, 
{\it JHEP} \textbf{0808} (2008), 035. %% [0805.3898 [hep-th]].
%
%
%M. M. Roberts and S. A. Hartnoll, 
%\emph{  Pseudogap and time reversal breaking in a holographic superconductor,  }
%JHEP 0808 (2008) 035 [0805.3898 [hep-th]].
%https://doi.org/10.1088/1126-6708/2008/08/035. 
%
%%%%%%%%%%%%%%%%%%%%%%%%%%%%%%%%%%%%%%%%%%%%%%%%%%%
%%%%%%%%%%%%%%%%%%%%%%%%%%%%%%%%%%%%%%%%%%%%%%%%%%% 
%%%%%%%%%%%%%%%%%%%%%%%%%%%%%%%%%%%%%%%%%%%%%%%%%%%% 
%%%%%%%%%%%%%%%%%%%%%%%%%%%%%%%%%%%%%%%%%%%%%%%%%%%%
\bibitem{Hutasoit-Siopsis-Therrien2014}
J. A. Hutasoit, G. Siopsis, and J. Therrien, Conductivity of Strongly Coupled Striped Superconductor,  
%\emph{Conductivity of Strongly Coupled Striped Superconductor,}
{\it JHEP} {\bf 1401} (2014), 132.%% [1208.2964 [hep-th]]. 
%[1208.2964 [hep-th]].
%https://doi.org/10.1007/JHEP01(2014)132.  
%J. A. Hutasoit, G. Siopsis, and J. Therrien, 
%\emph{Conductivity of Strongly Coupled Striped Superconductor,}
%JHEP 1401 (2014) 132 [1208.2964 [hep-th]].
%https://doi.org/10.1007/JHEP01(2014)132.  

 
%%%%%%%%%%%%%%%%%%%%%%%%%%%%%%%%%%%%%%%%%%%%%%%%%%%
%%%%%%%%%%%%%%%%%%%%%%%%%%%%%%%%%%%%%%%%%%%%%%%%%%%
%%%%%%%%%%%%%%%%%%%%%%%%%%%%%%%%%%%%%%%%%%%%%%%%%%%
%%%%%%%%%%%%%%%%%%%%%%%%%%%%%%%%%%%%%%%%%%%%%%%%%%%
\bibitem{Landsteiner-Liu2015}
K.~Landsteiner and Y.~Liu, The holographic Weyl semi-metal, 
%\emph{The holographic Weyl semi-metal}, 
{\it Phys. Lett.\ B} {\bf 753} (2016), 453-457. % [1505.04772[hep-th]].
% 
%K. Landsteiner and Y. Liu, 
%\emph{The holographic Weyl semi-metal}, 
%Phys. Lett.\ B {\bf 753} (2016) 453-457 [1505.04772[hep-th]].
%https://doi.org/10.1016/j.physletb.2015.12.052.
%%%%%%%%%%%%%%%%%%%%%%%%%%%%%%%%%%%%%%%%%%%%%%%%%%%
%%%%%%%%%%%%%%%%%%%%%%%%%%%%%%%%%%%%%%%%%%%%%%%%%%%
%%%%%%%%%%%%%%%%%%%%%%%%%%%%%%%%%%%%%%%%%%%%%%%%%%%
%%%%%%%%%%%%%%%%%%%%%%%%%%%%%%%%%%%%%%%%%%%%%%%%%%%


\bibitem{Landsteiner2015} 
K.~Landsteiner, Y.~Liu and Y.~M.~Sun, Quantum phase transition between a topological and a trivial semimetal from holography, 
%{\em Quantum phase transition between a topological and a trivial semimetal from holography,}
{\it Phys. Rev. Lett.}  {\bf 116} (2016), 081602. %%[1511.05505 [hep-th]].
 
% 
%  K.~Landsteiner, Y.~Liu and Y.~W.~Sun,
%{\em Quantum phase transition between a topological and a trivial semimetal from holography,}
%Phys. Rev. Lett.  {\bf 116} (2016) 081602 (2016)
%[1511.05505[hep-th]].
%https://doi.org/10.1103/PhysRevLett.116.081602.


%@article{PhysRevLett.116.081602,
%  title = {Quantum Phase Transition between a Topological and a Trivial Semimetal from Holography},
%  author = {Landsteiner, Karl and Liu, Yan and Sun, Ya-Wen},
%  journal = {Phys. Rev. Lett.},
%  volume = {116},
%  issue = {8},
%  pages = {081602},
%  numpages = {6},
%  year = {2016},
%  month = {Feb},
%  publisher = {American Physical Society},
%  doi = {10.1103/PhysRevLett.116.081602},
%  url = {https://link.aps.org/doi/10.1103/PhysRevLett.116.081602}
%}

%%%%%%%%%%%%%%%%%%%%%%%%%%%%%%%%%%%%%%%%%%%%%%%%%%%
%%%%%%%%%%%%%%%%%%%%%%%%%%%%%%%%%%%%%%%%%%%%%%%%%%%
%%%%%%%%%%%%%%%%%%%%%%%%%%%%%%%%%%%%%%%%%%%%%%%%%%%
%%%%%%%%%%%%%%%%%%%%%%%%%%%%%%%%%%%%%%%%%%%%%%%%%%
%%%%%%%%%%%%%%%%%%%%%%%%%%%%%%%%%%%%%%%%%%%%%%%%%%
\bibitem{Liu-Sun2018TI}
Y.~Liu and Y.~W.~Sun, Topological invariants for holographic semimetals, 
%\emph{ Topological invariants for holographic semimetals}
%  \doi{/}
{\it JHEP} {\bf 1810} (2018), 189.%% [1809.00513 [hep-th]]. 
%[1809.00513 [hep-th] ].  
%https://doi.org/10.1007/JHEP10(2018)189.
% 
%
%Y.~Liu and Y-W.~Sun,
%\emph{ Topological invariants for holographic semimetals}
%  \doi{/}
%JHEP {\bf 10} (2018) 189 [1809.00513 [hep-th] ].  
%https://doi.org/10.1007/JHEP10(2018)189.
%%%%%%%%%%%%%%%%%%%%%%%%%%%%%
%%%%%%%%%%%%%%%%%%%%%%%%%%%%%
%%%%%%%%%%%%%%%%%%%%%%%%%%%%%
\bibitem{Chernoff1973}
P.~R.~Chernoff, Essential self-adjointness of powers of generators of hyperbolic equations,
%\emph{Essential self-adjointness of powers of generators of hyperbolic equations}, 
{\it J. Funct. Anal.} \textbf{12}  (1973), 401-414. 
%
%P.R. Chernoff, 
%\emph{Essential self-adjointness of powers of generators of hyperbolic equations}, 
%J.Funct. Anal. \textbf{12} (1973) 401-414.
%https://doi.org/10.1016/0022-1236(73)90003-7.


%%%%%%%%%%%%%%%%%%%%%%%%%%%%%%%%%
%%%%%%%%%%%%%%%%%%%%%%%%%%%%%%%%%%%%%%%%%%%%%%%%%%%%
%%%%%%%%%%%%%%%%%%%%%%%%%%%%%%%%%%%%%%%%%%%%%%%%%%%
%\bibitem{Horowitz-Marolf1995}
\bibitem{Horowitz-Marolf1995}
G. T. Horowitz and D. Marolf,  Quantum probes of spacetime singularities,  
{\it Phys. Rev. }D \textbf{52} (1995), 5670-5675. % [9504028 [gr-qc]].
%G. T. Horowitz and D. Marolf, 
%\emph{  Quantum probes of spacetime singularities, }
%Phys. Rev. D \textbf{52} (1995) 5670-5675 [gr-qc/9504028].
%https://doi.org/10.1103/PhysRevD.52.5670.
%Quantum probes of spacetime singularities
%Phys. Rev. D \textbf{52} (1995) 5670, https://doi.org/10.1103/PhysRevD.52.5670.  
%
%G.~T.~Horowitz and D.~Marolf,
%Quantum probes of spacetime singularities
%Phys. Rev. D \textbf{52} (1995) 5670, https://doi.org/10.1103/PhysRevD.52.5670.  
%
%@article{PhysRevD.52.5670,
%  title = {Quantum probes of spacetime singularities},
%  author = {Horowitz, Gary T. and Marolf, Donald},
%  journal = {Phys. Rev. D},
%  volume = {52},
%  issue = {10},
%  pages = {5670--5675},
%  numpages = {0},
%  year = {1995},
%  month = {Nov},
%  publisher = {American Physical Society},
%  doi = {10.1103/PhysRevD.52.5670},
%  url = {https://link.aps.org/doi/10.1103/PhysRevD.52.5670}
%}
%%%%%%%%%%%%%%%%%%%%%%%%%%%%%%%%%%%%%%%%%%%%%%%%%%%
%%%%%%%%%%%%%%%%%%%%%%%%%%%%%%%%%%%%%%%%%%%%%%%%%%%
\bibitem{Astekar-Magnon1975}
A. Ashtekar and A. Magnon, Quantum fields in curved space-times, 
%\emph{Quantum fields in curved space-times},
{\it Proc. Roy. Soc. Lond.} A \textbf{346} (1975), 375-394.
%%
%A. Ashtekar and A. Magnon,
%\emph{Quantum fields in curved space-times},
%Proc. Roy. Soc. Lond. A \textbf{346} (1975) 375-394.
%https://doi.org/10.1098/rspa.1975.0181.

%%%%%%%%%%%%%%%%%%%%%%%%%%%%%%%%%%%%%%%%%%%%%%%%%%%
%%%%%%%%%%%%%%%%%%%%%%%%%%%%%%%%%%%%%%%%%%%%%%%%%%%
\bibitem{Hofmann-Schneider2015}
S. Hofmann and M. Schneider, Classical versus quantum completeness,  
%\emph{Classical versus quantum completeness},
{\it Phys. Rev.} D \textbf{91} (2015), 125028. % [1504.05580 [hep-th]].
%
%S. Hofmann and M. Schneider,
%\emph{Classical versus quantum completeness},
%Phys. Rev. D \textbf{91} (2015) 125028.
%https://doi.org/10.1103/PhysRevD.91.125028.
   %issue = {12},
  %pages = {125028
%@article{PhysRevD.91.125028,
 % title = {Classical versus quantum completeness},
  %author = {Hofmann, Stefan and Schneider, Marc},
  %journal = {Phys. Rev. D},
  %volume = {91},
  %issue = {12},
  %pages = {125028},
  %numpages = {6},
  %year = {2015},
  %month = {Jun},
  %publisher = {American Physical Society},
  %doi = {10.1103/PhysRevD.91.125028},
  %url = {https://link.aps.org/doi/10.1103/PhysRevD.91.125028}
%}

%%%%%%%%%%%%%%%%%%%%%%%%%%%%%%%%%%%%%%%%%%%%%%%%%%%%
%%%%%%%%%%%%%%%%%%%%%%%%%%%%%%%%%%%%%%%%%%%%%%%%%%%% 
%%%%%%%%%%%%%%%%%%%%%%%%%%%%%%%%%%%%%%%%%%%%%%%%%%%%
%%%%%%%%%%%%%%%%%%%%%%%%%%%%%%%%%%%%%%%%%%%%%%%%%%%% 
\bibitem{IT2019}
I. Tanaka, Quantum singularity in topological insulators,  
{\it J. Phys.: Condens. Matter. }\textbf{31} (2019), 255701. 
%%%%%%%%%%%%%%%%%%%%%%%%%%%%%%%
%I. Tanaka, 
%\emph{ Quantum singularity in topological insulators,  }
%J. Phys.: Condens. Matter. \textbf{31} (2019) 255701. 
%http://dx.doi.org/10.1088/1361-648X/ab1233.
%
%
%TY  - JOUR
%DO  - 10.1088/1361-648x/ab1233
%UR  - http://dx.doi.org/10.1088/1361-648X/ab1233
%TI  - Quantum singularity in topological insulators
%T2  - Journal of Physics: Condensed Matter
%AU  - Tanaka, Izumi
%PY  - 2019
%DA  - 2019/04/10
%PB  - IOP Publishing
%SP  - 255701
%IS  - 25
%VL  - 31
%SN  - 0953-8984
%SN  - 1361-648X
%AB  - Dislocations in the three-dimensional topological insulator have a similar effect as a hole threading magnetic flux. In this study, we have analytically considered a dislocation in three-dimensional topological insulator using the Dirac and the modified Dirac equation. A quantum singularity that originated in the dislocation creates a defect in momentum space. This defect relates to a phase transition and causes topologically protected zero-energy mode bound to the quantum singularity.
%ER  - 
%%%%%%%%%%%%%%%%%%%
%%%%%%%%%%%%%%%%%%%
%%%%%%%%%%%%%%%%%%%
\bibitem{Swingle2012} 
B.~Swingle, Entanglement renormalization and holography,
{\it Phys. Rev. }D \textbf{86} (2012), 065007. 

%@article{PhysRevD.86.065007,  title = {Entanglement renormalization and holography},  author = {Swingle, Brian},  journal = {Phys. Rev. D},  volume = {86},  issue = {6},  pages = {065007},  numpages = {8},  year = {2012},  month = {Sep},  publisher = {American Physical Society},  doi = {10.1103/PhysRevD.86.065007},  url = {https://link.aps.org/doi/10.1103/PhysRevD.86.065007}  

%%%%%%%%%%%%%%%%%%%%%%
%%%%%%%%%%%%%%%%%%%%%%
%%%%%%%%%%%%%%%%%%%%%%
\bibitem{Nozaki-Ryu-Takayanagi2012}
M.~Nozaki, S.~Ryu and T.~Takayanagi, Holographic geometry of entanglement renormalization in quantum field theories,
{\it JHEP }\textbf{10} (2012), 193. 
 %https://doi.org/10.1007/JHEP10(2012)193 


%%%%%%%%%%%%%%%%%%%%%%
%%%%%%%%%%%%%%%%%%%%%%
%%%%%%%%%%%%%%%%%%%%%%
\bibitem{ZLSS2015}
J.~Zaanen, Y.~Liu, Y-W.~Sun, and K.~Schalm, 
\textit{Holographic Duality in Condensed Matter Physics}
% (2015). Holographic Duality in Condensed Matter Physics. 
 (Cambridge University Press, Cambridge, 2015).
 %doi:10.1017/CBO9781139942492 

%%%%%%%%%%%%%%%%%%%%%%
%%%%%%%%%%%%%%%%%%%%%%
%%%%%%%%%%%%%%%%%%%%%%
\bibitem{AFZLS2014}
D. Are\'an, A. Farahi, L. A. P. Zayas, I. S. Landea, and A. Scardicchio,  
Holographic superconductor with disorder, 
{\it Phys. Rev. D} \textbf{89} (2014), 106003. 
%@article{PhysRevD.89.106003,
%  title = {Holographic superconductor with disorder},
%  author = {Are\'an, D. and Farahi, A. and Pando Zayas, L. A. and Landea, I. Salazar and Scardicchio, A.},
%  journal = {Phys. Rev. D},
 % volume = {89},
%  issue = {10},
%  pages = {106003},
%  numpages = {6},
%  year = {2014},
%  month = {May},
%  publisher = {American Physical Society},
%  doi = {10.1103/PhysRevD.89.106003},
%  url = {https://link.aps.org/doi/10.1103/PhysRevD.89.106003}
%}
%%%%%%%%%%%%%%%%%%%%%%%%%%%%%%%%%%%%%%%%%%%%%%%%%%%%
%%%%%%%%%
%%%%%%%%%%%%%%%%%%%%%%%%%%%%%%%%%%%%%%%%%%%%%
%%%%%%%%%%%%%%%%%%%%%%%%%%%%%%%%%%%%%%%%%%%%
 \bibitem{Koga-Maeda-Tomoda2014}
J.~Koga, K.~Maeda, and K.~Tomoda, 
Holographic superconductor model in a spatially anisotropic background, 
{\it Phys. Rev. D} \textbf{89} (2014), 104024. 

%\UTF{2013} Published 14 May 2014 
% @article{PhysRevD.89.104024,
%  title = {Holographic superconductor model in a spatially anisotropic background},
%  author = {Koga, Jun-ichirou and Maeda, Kengo and Tomoda, Kentaro},
%  journal = {Phys. Rev. D},
%  volume = {89},
%  issue = {10},
%  pages = {104024},
%  numpages = {9},
%  year = {2014},
%  month = {May},
%  publisher = {American Physical Society},
%  doi = {10.1103/PhysRevD.89.104024},
%  url = {https://link.aps.org/doi/10.1103/PhysRevD.89.104024}
%}
%%%%%%%%%%%%%%%%%%%%%%%%%%%%%%%%%%%%%%%%%%%%%%%%%%%%
%%%%%%%%%
%%%%%%%%%%%%%%%%%%%%%%%%%%%%%%%%%%%%%%%%%%%%%
%%%%%%%%%%%%%%%%%%%%%%%%%%%%%%%%%%%%%%%%%%%% 
\bibitem{Donos-Hartnoll2013}
A.~Donos, S.~Hartnoll, Interaction-driven localization in holography, 
{\it Nature Phys }\textbf{9} (2013), 649-655. 
%https://doi.org/10.1038/nphys2701 
%Interaction-driven localization in holography

%%%%%%%%%%%%%%%%%%%%%%%%%%%%%%%%%%%%%%%%%%%%%%%%%%%%
%%%%%%%%%
%%%%%%%%%%%%%%%%%
%%%%%%%%%%%%%%%%%
\bibitem{Donos-Gauntlett-Pantelidou2015}
A. Donos, J.P. Gauntlett and C. Pantelidou, Conformal field theories in $d=4$ with a helical twist, 
%\emph{Conformal field theories in $d=4$ with a helical twist}, 
{\it Phys. Rev. D.} \textbf{91} (2015), 066003. %% [1412.3446 [hep-th]].

%A. Donos, J.P. Gauntlett and C. Pantelidou, 
%\emph{Conformal field theories in $d=4$ with a helical twist}, 
%Phys. Rev. D. \textbf{91} (2015) 066003 [1412.3446 [hep-th]].
%https://doi.org/10.1103/PhysRevD.91.066003 

%
%@article{PhysRevD.91.066003,
%  title = {Conformal field theories in $d=4$ with a helical twist},
%  author = {Donos, Aristomenis and Gauntlett, Jerome P. and Pantelidou, Christiana},
%  journal = {Phys. Rev. D},
%  volume = {91},
%  issue = {6},
%  pages = {066003},
%  numpages = {18},
%  year = {2015},
%  month = {Mar},
%  publisher = {American Physical Society},
%  doi = {10.1103/PhysRevD.91.066003},
%  url = {https://link.aps.org/doi/10.1103/PhysRevD.91.066003}
%}
%%%%%%%%%%%%%%%%%% 
%%%%%%%%%%%%%%%%%%%%%%%%%%%%%%%%%%%%%%%%%%%%%
%%%%%%%%%%%%%%%%%%%%%%%%%%%%%%%%%%%%%%%%%%%%\bibitem{}
\bibitem{RGT2011}
C. de Rham, G.~Gabadadze, and A.~J. Tolley, 
Resummation of Massive Gravity, 
{\it Phys. Rev. Lett.} \textbf{106} (2011), 231101. 

%\UTF{2013} Published 10 June 2011
%@article{PhysRevLett.106.231101,
 % title = {Resummation of Massive Gravity},
%  author = {de Rham, Claudia and Gabadadze, Gregory and Tolley, Andrew J.},
 % journal = {Phys. Rev. Lett.},
 % volume = {106},
 % issue = {23},
%  pages = {231101},
%  numpages = {4},
%  year = {2011},
%  month = {Jun},
%  publisher = {American Physical Society},
%  doi = {10.1103/PhysRevLett.106.231101},
%  url = {https://link.aps.org/doi/10.1103/PhysRevLett.106.231101}
%}
%%%%%%%%%%%%%%%%%%%%%%%%%%%%%%%%%%%%%%%%%%%%%%%%%%%%
%%%%%%%%%
%%%%%%%%%%%%%%%%%%%%%%%%%%%%%%%%%%%%%%%%%%%%%
%%%%%%%%%%%%%%%%%%%%%%%%%%%%%%%%%%%%%%%%%%%%
\bibitem{NOP2010}
S.~Nakamura, H.~Ooguri, and C-S.~Park, Gravity dual of spatially modulated phase,  
{\it Phys. Rev. D} \textbf{81} (2010), 044018.

% \UTF{2013} Published 10 February 2010
% @article{PhysRevD.81.044018,  title = {Gravity dual of spatially modulated phase},  author = {Nakamura, Shin and Ooguri, Hirosi and Park, Chang-Soon},  journal = {Phys. Rev. D},  volume = {81},  issue = {4},  pages = {044018},  numpages = {12}, year = {2010},  month = {Feb},  publisher = {American Physical Society},  doi = {10.1103/PhysRevD.81.044018},  url = {https://link.aps.org/doi/10.1103/PhysRevD.81.044018}}


%%%%%%%%%%%%%%%%%%%%%%%%%%%%%%%%%%%%%%%%%%%%%%%%%%%%
%%%%%%%%%
%%%%%%%%%%%%%%%%%%%%%%%%%%%%%%%%%%%%%%%%%%%%%
%%%%%%%%%%%%%%%%%%%%%%%%%%%%%%%%%%%%%%%%%%%%

\bibitem{Ooguri-Park2010}
H.~Ooguri and C-S.~Park, Holographic endpoint of spatially modulated phase transition,  
{\it Phys. Rev. D} \textbf{82} (2010), 126001. 
%\UTF{2013} Published 2 December 2010
%@article{PhysRevD.82.126001,  title = {Holographic endpoint of spatially modulated phase transition},  author = {Ooguri, Hirosi and Park, Chang-Soon},  journal = {Phys. Rev. D},  volume = {82},  issue = {12},  pages = {126001},  numpages = {10},  year = {2010}, month = {Dec},  publisher = {American Physical Society},  doi = {10.1103/PhysRevD.82.126001},  url = {https://link.aps.org/doi/10.1103/PhysRevD.82.126001}}

%%%%%%%%%%%%%%%%%%%%%%%%%%%%%%%%%%%%%%%%%%%%%%%%%%%%
%%%%%%%%%
%%%%%%%%%%%%%%%%%%%%%%%%%%%%%%%%%%%%%%%%%%%%%
%%%%%%%%%%%%%%%%%%%%%%%%%%%%%%%%%%%%%%%%%%%%
\bibitem{HKT2012}
S.~Harrison, S.~Kachru and G.~ Torroba, A maximally supersymmetric Kondo model, 
{\it Class. Quantum Grav.} \textbf{29} (2012), 194005. 

%%%%%%%%%%%%%%%%%%%%%%%%%%%%%%%%%%%%%%%%%%%%%%%%%%%%
%%%%%%%%%
%%%%%%%%%%%%%%%%%%%%%%%%%%%%%%%%%%%%%%%%%%%%%
%%%%%%%%%%%%%%%%%%%%%%%%%%%%%%%%%%%%%%%%%%%%
\bibitem{EHOJ2013} 
J.~Erdmenger, C.~Hoyos, A.~O’Bannon, and W.~Jackson,
A holographic model of the Kondo effect, {\it JHEP} \textbf{12} (2010), 086.
%https://doi.org/10.1007/JHEP12(2013) 086

%%%%%%%%%%%%%%%%%%%%%%%%%%%%%%%%%%%%%%%%%%%%%%%%%%%%
%%%%%%%%%
%%%%%%%%%%%%%%%%%%%%%%%%%%%%%%%%%%%%%%%%%%%%%
%%%%%%%%%%%%%%%%%%%%%%%%%%%%%%%%%%%%%%%%%%%%
\bibitem{KKY2010}
S.~Kachru, A.~Karch and S.~Yaida, 
Adventures in holographic dimer models, 
{\it New Journal of Physics} \textbf{13} (2011), 035004.

%%%%%%%%%%%%%%%%%%%%%%%%%%%%%%%%%%%%%%%%%%%%%%%%%%%%
%%%%%%%%%
%%%%%%%%%%%%%%%%%%%%%%%%%%%%%%%%%%%%%%%%%%%%%
%%%%%%%%%%%%%%%%%%%%%%%%%%%%%%%%%%%%%%%%%%%%
%%%%%%%%%%%%%%%%%%%%%%%%%%%%%%%%%%%%%%%%%%%%%%%%%%%
%%%%%%%%%%%%%%%%%%%%%%%%%%%%%%%%%%%%%%%%%%%%%%%%%%%
%%%%%%%%%%%%%%%%%%%%%%%%%%%%%%%%%%%%%%%%%%%%%%%%%%%% 
\bibitem{Shen2012} 
S-Q Shen,  \textit{Topological Insulators} (Springer, Berlin, 2012). 
%%%%%%%%%%%%%%%%%%%%%%%%%%%%%%%%%%%%%%%%%%%%%%%%%%%
%%%%%%%%%%%%%%%%%%%%%%%%%%%%%%%%%%%%%%%%%%%%%%%%%%%
%%%%%%%%%%%%%%%%%%%%%%%%%%%%%%%%%%%%%%%%%%%%%%%%%%%% 　
\bibitem{Chesler-Hong-Adams}
P.~M.~Chesler, L.~H.~Hong and A.~Adams, 
Holographic Vortex Liquids and Superfluid Turbulence, 
{\it Science} \textbf{341} (2013), 368-372. %% [1212.0281 [hep-th]].  
%%%%%%%%%%%%%%%%%%%%%%%%%%%%%%%%%%%%%%%%%%%%%%%%%%%
%%%%%%%%%%%%%%%%%%%%%%%%%%%%%%%%%%%%%%%%%%%%%%%%%%%
%%%%%%%%%%%%%%%%%%%%%%%%%%%%%%%%%%%%%%%%%%%%%%%%%%%% 

\bibitem{Galtsov-Letelier} %PRD 47 4273  
D. V. Gal'tsov and P.S. Letelier, 
Spinning strings and cosmic dislocations,   
{\it Phys. Rev. }D\textbf{47} (1993), 4273-4276.
%%%%
%D. V. Gal'tsov and P.S. Letelier, 
%\emph{Spinning strings and cosmic dislocations},    
%Phys. Rev. D\textbf{47} (1993) 4273-4276.
%https://doi.org/10.1103/PhysRevD.47.4273.
%
%@article{PhysRevD.47.4273,
%  title = {Spinning strings and cosmic dislocations},
%  author = {Gal'tsov, D. V. and Letelier, P. S.},
%  journal = {Phys. Rev. D},
% volume = {47},
%  issue = {10},
%  pages = {4273--4276},
%  numpages = {0},
%  year = {1993},
%  month = {May},
%  publisher = {American Physical Society},
%  doi = {10.1103/PhysRevD.47.4273},
%  url = {https://link.aps.org/doi/10.1103/PhysRevD.47.4273}
%}

%%%%%%%%%%%%%%%%%%%%%%%%%%%%%%%%%%%%%%%%%%%%%%%%%%%
%%%%%%%%%%%%%%%%%%%%%%%%%%%%%%%%%%%%%%%%%%%%%%%%%%%
\bibitem{Furtado-Berezza-Moraes2001}
C. Furtado, V. B. Bezerra and F. Moraes, Quantum scattering by a magnetic flux screw dislocation, 
%\emph{Quantum scattering by a magnetic flux screw dislocation},
{\it Phys. Lett.} A\textbf{289} (2001), 160-166.
%
%
%C. Furtado, V. B. Bezerra and F. Moraes, 
%\emph{Quantum scattering by a magnetic flux screw dislocation},
%Phys. Lett. A\textbf{289} (2001) 160-166.
%https://doi.org/10.1016/S0375-9601(01)00615-6. 
%
%Physics Letters A
%Volume 289, Issue 3, 15 October 2001, Pages 160-166	?
%Quantum scattering by a magnetic flux screw dislocation
%Author links open overlay panelCl?udioFurtadoaV.BBezerrabFernandoMoraesa
%Show more
%https://doi.org/10.1016/S0375-9601(01)00615-6

%%%%%%%%%%%%%%%%%%%%%%%%%%%%%%%%%%%%%%%%%%%%%%%%%%%
%%%%%%%%%%%%%%%%%%%%%%%%%%%%%%%%%%%%%%%%%%%%%%%%%%%
%%%%%%%%%%%%%%%%%%%%%%%%%%%%%%%%%%%%%%%%%%%%%%%%%%%
\bibitem{Chen-Wang-Su2006} 
S.~Chen, B.~Wang and R.~Su, 
Influence of Lorentz violation on Dirac quasinormal modes in the Schwarzschild black hole spacetime,
%\emph{Influence of Lorentz violation on Dirac quasinormal modes in the Schwarzschild black hole spacetime}, 
{\it Class. Quantum Grav.} \textbf{23} (2006), 7581-7590. %% [0701089 [gr-qc]].
%https://doi.org/10.1088/0264-9381/23/24/026.  
%
%
%S.~Chen, B.~Wang and R.~Su,
%\emph{Influence of Lorentz violation on Dirac quasinormal modes in the Schwarzschild black hole spacetime}, 
%Class. Quantum Grav. \textbf{23} (2006) 7581.
%https://doi.org/10.1088/0264-9381/23/24/026. 




%%%%%%%%%%%%%%%%%%%%%%%%%%%%%%%%%%%%%%%%%%%%%%%%%%%%%
\bibitem{Pallab-Sumanta2013}
G.~Pallab and T.~Sumanta, Axionic field theory of $(3+1)$-dimensional Weyl semimetals, 
%\emph{Axionic field theory of $(3+1)$-dimensional Weyl semimetals},
{\it Phys. Rev. }B \textbf{88} (2013), 245107. %% [1210.6352 [cond-mat.mes-hall]]. 
%arXiv:1210.6352 [cond-mat.mes-hall]. 
%G.~Pallab and T.~Sumanta,
%\emph{Axionic field theory of $(3+1)$-dimensional Weyl semimetals},
%Phys. Rev. B \textbf{88} (2013) 245107 arXiv:1210.6352 [cond-mat.mes-hall].
%https://doi.org/10.1103/PhysRevB.88.245107. 
%@article{PhysRevB.88.245107,
%  title = {Axionic field theory of $(3+1)$-dimensional Weyl semimetals},
%  author = {Goswami, Pallab and Tewari, Sumanta},
%  journal = {Phys. Rev. B},
%  volume = {88},
%  issue = {24},
%  pages = {245107},
%  numpages = {9},
%  year = {2013},
%  month = {Dec},
%  publisher = {American Physical Society},
%  doi = {10.1103/PhysRevB.88.245107},
%  url = {https://link.aps.org/doi/10.1103/PhysRevB.88.245107}
%}

%%%%%%%%%%%%%%%%%%%%%%%%%%%%%%%%%%%%%%%%%%%%%%%%%%%%%
%%%%%%%%%%%%%%%%%%%%%%%%%%%%%%%%%%%%%%%%%%%%%%%%%%%%%
%%%%%%%%%%%%%%%%%
%%%%%%%%%%%%%%%%%
\bibitem{stephanov-yin2012}
M.~A.~Stephanov and Y.~Yin, Chiral kinetic theory,  
%\emph{Chiral kinetic theory}, 
{\it Phys. Rev. Lett. }\textbf{109} (2012), 162001. % [1207.0747 [hep-th]]. 
%[1207.0747 [hep-th]].  
%
%
%M.~A.~Stephanov and Y.~Yin, 
%\emph{Chiral kinetic theory}, 
%Phys. Rev. Lett. \textbf{109} (2012) 162001 [1207.0747 [hep-th]]. 
%https://doi.org/10.1103/PhysRevLett.109.162001. 
%@article{PhysRevLett.109.162001,
%  title = {Chiral Kinetic Theory},
%  author = {Stephanov, M. A. and Yin, Y.},
%  journal = {Phys. Rev. Lett.},
%  volume = {109},
%  issue = {16},
%  pages = {162001},
%  numpages = {5},
%  year = {2012},
%  month = {Oct},
%  publisher = {American Physical Society},
%  doi = {10.1103/PhysRevLett.109.162001},
%  url = {https://link.aps.org/doi/10.1103/PhysRevLett.109.162001}
%arXiv:1207.0747v1 [hep-th] 



%%%%%%%%%%%%%%%%%%%%
%%%%%%%%%%%%%%%%%%%%
\bibitem{stone-dwivedi2013} 
M.~Stone and V.~Dwivedi, Classical version of the non-abelian gauge anomaly, 
%\emph{Classical version of the non-abelian gauge anomaly}, 
{\it Phys. Rev. D.} \textbf{88} (2013), 045012. % [1305.1955 [hep-th]]. 
%M.~Stone and V.~Dwivedi,
%\emph{Classical version of the non-abelian gauge anomaly}, 
%Phys. Rev. D. \textbf{88} (2013) 045012 [1305.1955 [hep-th]]. 
%https://doi.org/10.1103/PhysRevD.88.045012. 
%
% @article{PhysRevD.88.045012,
%  title = {Classical version of the non-Abelian gauge anomaly},
%  author = {Stone, Michael and Dwivedi, Vatsal},
%  journal = {Phys. Rev. D},
%  volume = {88},
%  issue = {4},
%  pages = {045012},
%  numpages = {8},
%  year = {2013},
%  month = {Aug},
%  publisher = {American Physical Society},
%  doi = {10.1103/PhysRevD.88.045012},
%  url = {https://link.aps.org/doi/10.1103/PhysRevD.88.045012}
%arXiv:1305.1955 [hep-th]
   
%%%%%%%%%%%%%%%%%%%%%%%%%%
%%%%%%%%%%%%%%%%%%%%%%%%%%   
\bibitem{Ishihara-Mizushima-Tsuruta-Fujimoto2019}
Y.~Ishihara, T.~Mizushima, A.~ Tsuruta and S.~Fujimoto, 
Torsional chiral magnetic effect due to skyrmion textures in a Weyl superfluid $^{3}\mathrm{He}\text{\ensuremath{-}}\mathrm{A}$, 
%\emph {Torsional chiral magnetic effect due to skyrmion textures in a Weyl superfluid $^{3}\mathrm{He}\text{\ensuremath{-}}\mathrm{A}$}, 
{\it Phys. Rev. }B. \textbf{99} (2019), 025413. % [1807.09994 [cond-mat.supr-con]].   
%
%
%Y.~Ishihara, T.~Mizushima, A.~ Tsuruta and S.~Fujimoto
%\emph {Torsional chiral magnetic effect due to skyrmion textures in a Weyl superfluid $^{3}\mathrm{He}\text{\ensuremath{-}}\mathrm{A}$}, 
%Phys. Rev. B. \textbf{99} (2019) 025413 [1807.09994 [cond-mat.supr-con]]. 
%https://doi.org/10.1103/PhysRevB.99.024513. 
  % 
%%@article{PhysRevB.99.024513,
%  title = {Torsional chiral magnetic effect due to skyrmion textures in a Weyl superfluid $^{3}\mathrm{He}\text{\ensuremath{-}}\mathrm{A}$},
%  author = {Ishihara, Yusuke and Mizushima, Takeshi and Tsuruta, %Atsushi and Fujimoto, Satoshi},
%  journal = {Phys. Rev. B},
%  volume = {99},
%  issue = {2},
%  pages = {024513},
%  numpages = {13},
%  year = {2019},
%  month = {Jan},
%  publisher = {American Physical Society},
%  doi = {10.1103/PhysRevB.99.024513},
%  url = {https://link.aps.org/doi/10.1103/PhysRevB.99.024513}
%}


%%%%%%%%%%%%%%%%%%%%%%%%%
%%%%%%%%%%%%%%%%%%%%%%%%%
\bibitem{Marques-Furtado-Bezerra-Fernando2001}
G.~A.~Marques, C.~Furtado, V.~B.~Bezerra, and F.~Moraes, 
Landau levels in the presence of topological defects, 
%\emph{Landau levels in the presence of topological defects}, 
{\it Jour of Phys A: Mathematical and General.} \textbf{34} (2001), 5945-5954. % [0012146 [quant-ph]].
%G.~A.~Marques, C.~Furtado, V.~B.~Bezerra, and F.~Moraes, 
%\emph{Landau levels in the presence of topological defects}, 
%Jour of Phys A: Mathematical and General. \textbf{34} (2001) 5945 [0012146 [quant-ph]].
%http://dx.doi.org/10.1088/0305-4470/34/30/306. 
% UR  - http://dx.doi.org/10.1088/0305-4470/34/30/306
%TI  - Landau levels in the presence of topological defects
%T2  - Journal of Physics A: Mathematical and General
%AU  - Marques, Geusa de A
%AU  - Furtado, Claudio
%AU  - Bezerra, V B
%AU  - Moraes, Fernando
%PY  - 2001
%DA  - 2001/07/20
%PB  - IOP Publishing
%SP  - 5945-5954
%IS  - 30
%VL  - 34
%%%%%%%%%%%%%%%%%%%%%%%%%%%%%%%%%%%%%%%%%%%%%%%%%%%
%%%%%%%%%%%%%%%%%%%%%%%%%%%%%%%%%%%%%%%%%%%%%%%%%%%

%%%%%%%%%%%%%%%%%%%%%%%%%%%%%%%%%%%%%%%%%%%%%%%%%%%
%%%%%%%%%%%%%%%%%%%%%%%%%%%%%%%%%%%%%%%%%%%%%%%%%%%
%\cite{Iqbal:2009fd}
\bibitem{Iqbal:2009fd} 
N.~Iqbal and H.~Liu, Real-time response in AdS/CFT with application to spinors, 
%\emph{ Real-time response in AdS/CFT with application to spinors,}
%  \doi{10.1002/prop.200900057}
{\it Fortsch.\ Phys.\ }  {\bf 57} (2009), 367-384. % [0903.2596 [hep-th]].  
%-384 [0903.2596 [hep-th]].   

%  N.~Iqbal and H.~Liu,
%\emph{ Real-time response in AdS/CFT with application to spinors,}
%  \doi{10.1002/prop.200900057}
%Fortsch.\ Phys.\  {\bf 57} (2009) 367-384 [0903.2596 [hep-th]].  
%  [\arXiv{0903.2596}{hep-th}].
  %%CITATION = doi:10.1002/prop.200900057;%%
%https://doi.org/10.1002/prop.200900057. 

%%%%%%%%%%%%%%%%%%%%%%%%%%%%%%%%%%%%%%%%%%%%%%%%%%%
%%%%%%%%%%%%%%%%%%%%%%%%%%%%%%%%%%%%%%%%%%%%%%%%%%%%%%%%
%%%%%%%%%%%%%%%%%%%%%%%%%%%%%%%%%%%%%%%%%%%%%%%%%%%%%%%%

\bibitem{Oliveria-Tiomno1962}
C. G. de Oliveira and J. Tiomno, Representations of Dirac equation in general relativity,
%\emph{ Representations of Dirac equation in general relativity,   }
{\it J. Nuovo Cim.}  \textbf{24} (1962), 672. 
%https://doi.org/10.1007/BF02816716.
%C. G. de Oliveira and J. Tiomno, 
%\emph{ Representations of Dirac equation in general relativity,   }
%J. Nuovo Cim.  \textbf{24} (1962) 672. 
%https://doi.org/10.1007/BF02816716.
%TY  - JOUR
%AU  - de Oliveira, C. G.
%AU  - Tiomno, J.
%PY  - 1962
%DA  - 1962/05/01
%TI  - 
%Representations of Dirac equation in general relativity
%JO  - Il Nuovo Cimento (1955-1965)
%SP  - 672
%EP  - 687
%VL  - 24
%IS  - 4
%AB  - After a reformulation of the General Relativistic equations of motion of a point mass in a gravitational field in terms of \UTF{00AB} fourlegs \UTF{00BB}, the generalized Dirac equation is written in the Schr\UTF{00F6}dinger representation and the equations of motion of classical observable quantities compared with the previous one. The Foldy-Wouthuysen representation is obtained in presence of gravitational fields. The most interesting result is that the gravitational \UTF{00AB} gyromagnetic \UTF{00BB} factor is 1 instead of 2 as the electromagnetic one. The interpretation of this fact is that the spinor field (and probably all fields) describes rotating particles where the gravitational mass has exactly the same space distribution as the inertialo mass. This is not true in Moshinsky-Birkhoff linear theory. Finally the red shift of hydrogen atoms levels in presence of gravitational field is obtained and found to coincide with the usual prediction of General Relativity.
%SN  - 1827-6121
%UR  - https://doi.org/10.1007/BF02816716
%DO  - 10.1007/BF02816716
%ID  - de Oliveira1962
%ER  - 
%%%%%%%%%%%%%%%%%
%%%%%%%%%%%%%%%%%%%%%%%%%%%%%%%%%%%%%%%%%%%%%%%%%%%
%%%%%%%%%%%%%%%%%%%%%%%%%%%%%%%%%%%%%%%%%%%%%%%%%%% 
\bibitem{Ishibashi-Hosoya1999} 
A.~Ishibashi and A.~Hosoya, Who's afraid of naked singularities? Probing timelike singularities with finite energy waves, 
{\it Phys. Rev. } D \textbf{60} (1999), 104028. % [9907009 [gr-qc]]. 
% @article{PhysRevD.60.104028,
%  title = {Who's afraid of naked singularities? Probing timelike singularities with finite energy waves},
%  author = {Ishibashi, Akihiro and Hosoya, Akio},
%  journal = {Phys. Rev. D},
%  volume = {60},
%  issue = {10},
%  pages = {104028},
%  numpages = {12},
%  year = {1999},
%  month = {Oct},
%  publisher = {American Physical Society},
%  doi = {10.1103/PhysRevD.60.104028},
%  url = {https://link.aps.org/doi/10.1103/PhysRevD.60.104028}
%} 
 


%%%%%%%%%%%%%%%%%%%%%%%%%%%%%%%%%%%%%%%%%%%%%%%%%%%
%%%%%%%%%%%%%%%%%%%%%%%%%%%%%%%%%%%%%%%%%%%%%%%%%%% 
%%%%%%%%%%%%%%%%%%%%%%%%%%%%%%%%%%%%%%%%%%%%%%%%%%%
%%%%%%%%%%%%%%%%%%%%%%%%%%%%%%%%%%%%%%%%%%%%%%%%%%%
\bibitem{Tong_GT} 
D.~Tong,  %  2018 %: 
Lectures on Gauge Theory, (2018),
\url{http://www.damtp.cam.ac.uk/user/tong/gaugetheory/gt.pdf }

%%%%%%%%%%%%%%%%%
%%%%%%%%%%%%%%%%%%%%%%%%%%%%%%%%%%%%%%%%%%%%%%%%%%%
%%%%%%%%%%%%%%%%%%
%%%%%%%%%%%%%%%%%%%%%%%%%%%%%%%%%%%%%%%%%%%%%%%%%%%
%%%%%%%%%%%%%%%%%%%%%%%%%%%%%%%%%%%%%%%%%%%%%%%%%%% 
\bibitem{Hill2006}
C. T. Hill, Anomalies, Chern-Simons terms and chiral delocalization in extra dimensions, 
%\emph{Anomalies, Chern-Simons terms and chiral delocalization in extra dimensions}, 
{\it Phys. Rev.} D\textbf{73} (2006), 085001. % [0601154 [hep-th]].
%
% 
%\emph{Anomalies, Chern-Simons terms and chiral delocalization in extra dimensions}, 
%Phys. Rev. D\textbf{73} (2006) 085001 [hep-th/0601154].
%https://doi.org/10.1103/PhysRevD.73.085001.
%@article{PhysRevD.73.085001,
%  title = {Anomalies, Chern-Simons terms and chiral delocalization in extra dimensions},
%  author = {Hill, Christopher T.},
%  journal = {Phys. Rev. D},
%  volume = {73},
%  issue = {8},
%  pages = {085001},
%  numpages = {25},
%  year = {2006},
%  month = {Apr},
%  publisher = {American Physical Society},
%  doi = {10.1103/PhysRevD.73.085001},
%  url = {https://link.aps.org/doi/10.1103/PhysRevD.73.085001}
%}
%%%%%%%%%%%%%%%%%%%%%%%%%%%%%%%%%%%%%%%%%%%%%%%%%%%
%%%%%%%%%%%%%%%%%%%%%%%%%%%%%%%%%%%%%%%%%%%%%%%%%%%
%%%%%%%%%%%%%%%%%%%%%%%%%%%%%%%%%%
%%%%%%%%%%%%%%%%%%%%%%%%%%%%%%%%%%%%%%%%%%%%%%%%%%% 
\bibitem{Hughes et al 2016}
T.~L.~Hughes , R.~G.~Leigh G, O.~Parrikar and S.~T.~Ramamurthy, 
Entanglement entropy and anomaly inflow,  
{\it Phys. Rev.} D \textbf{93} (2016), 065059. %% [1509.04969 [hep-th]]. 
%@article{PhysRevD.93.065059,
%  title = {Entanglement entropy and anomaly inflow},
%  author = {Hughes, Taylor L. and Leigh, Robert G. and Parrikar, Onkar and Ramamurthy, Srinidhi T.},
%  journal = {Phys. Rev. D},
%  volume = {93},
%  issue = {6},
%  pages = {065059},
%  numpages = {18},
%  year = {2016},
%  month = {Mar},
%  publisher = {American Physical Society},
%  doi = {10.1103/PhysRevD.93.065059},
%  url = {https://link.aps.org/doi/10.1103/PhysRevD.93.065059}
%
%%%%%%%%%%%%%%%%%%%%%%%%%%%%%%%%%%%%%%%%%%%%%%%%%%%
%%%%%%%%%%%%%%%%%%%%%%%%%%%%%%%%%%
%%%%%%%%%%%%%%%%%%%%%%%%%%%%%%%%%%%%%%%%%%%%%%%%%%% 
%%%%%%%%%%%%%%%%%%%%%%%%%%%%%%%%%%%%%%%%%%%%%%%%%%%
%%%%%%%%%%%%%%%%%%%%%%%%%%%%%%%%%%
%%%%%%%%%%%%%%%%%%%%%%%%%%%%%%%%%%%%%%%%%%%%%%%%%%%
\bibitem{Gripaios-West2008}
B. Gripaios and S. M. West, Anomaly holography,
{\it Nucl. Phys.} B \textbf{789} (2008), 362-381.%% [0704.3981 [hep-ph]].


%B. Gripaios and S. M. West, \emph{ Anomaly holography },
%Nucl. Phys. B \textbf{789} (2008) 362-381 [0704.3981 [hep-ph]].
%https://doi.org/10.1016/j.nuclphysb.2007.08.008.  
%%%%%%%%%%%%%%%%%%%%%%%%%
%%%%%%%%%%%%%%%%%%%%%%%%%
%%%%%%%%%%%%%%%%%%%%%%%%%
\bibitem{Benini2018}   
F. ~Benini, Brief Introduction to AdS/CFT, (2018),
\url{https://www.sissa.it/tpp/phdsection/OnlineResources/16/SISSA_AdS_CFT_course2018.pdf}

%%%%%%%%%%%%%%%%%%%%%%%%%%%%%%%%%%%%%%%%%%%%%%%%%%
%%%%%%%%%%%%%%%%%%%%%%%%%%%%%%%%%%%%%%%%%%%%%%%%%%

%%%%%%%%%%%%%%%%%%%%%%%%%%%%%%%%%%%%%%%%%%%%%%%%%%%
%%%%%%%%%%%%%%%%%%%%%%%%%%%%%%%%%%%%%%%%%%%%%%%%%%%%
%%%%%
\bibitem{Witten-Yonekura2019}   
E.~Witten and K.~Yonekura,	 Anomaly Inflow and the $\eta$-Invariant, 
[1909.08775 [hep-th]].

%%%%%%%%%%%%%%%%%%%%%%%%%%%%%%%%%%%%%%%%%%%%%%%%%%%
%%%%%%%%%%%%%%%%%%%%%%%%%%%%%%%%%%%%%%%%%%%%%%%%%%%%
%%%%% 
\bibitem{Hatsugai2010} 
Y.~Hatsugai, 
Symmetry-protected-$\mathbb{Z}_2$-quantization and quaternionic Berry connection with Kramers degeneracy,
{\it New J. Phys.} \textbf{12} (2010), 065004. % [0909.4831 [cond-mat.mes-hall]].
  
%%%%%%%%%%%%%%%%%%%%%%%%%%%%%%%%%%%%%%%%%%%%%%%%%%%
%%%%%%%%%%%%%%%%%%%%%%%%%%%%%%%%%%%%%%%%%%%%%%%%%%%%
%%%%% 
 
\bibitem{Liu-McGreevy-Vegh2011}
H.~Liu, J.~McGreevy, and D.~Vegh, Non-Fermi liquids from holography,
{\it Phys. Rev.} D \textbf{83} (2011), 065029. % [0903.2477 [hep-th]].
%@article{PhysRevD.83.065029,
%  title = {Non-Fermi liquids from holography},
%  author = {Liu, Hong and McGreevy, John and Vegh, David},
%  journal = {Phys. Rev. D},
%  volume = {83},
%  issue = {6},
%  pages = {065029},
%  numpages = {10},
%  year = {2011},
%  month = {Mar},
%  publisher = {American Physical Society},
%  doi = {10.1103/PhysRevD.83.065029},
%  url = {https://link.aps.org/doi/10.1103/PhysRevD.83.065029}
% }

%%%%%%%%%%%%%%%%%%%%%%%
%%%%%%%%%%%%%%%%%%%%%%%
%%%%%%%%%%%%%%%%%%%%%%%
\bibitem{WCLGQR2016}
X.~Wen, G.~Y.~Cho, P.~L.~S. Lopes, Y.~Gu, X-L.~Qi, and S.~Ryu, 
Holographic entanglement renormalization of topological insulators, 
{\it Phys. Rev.} B\textbf{94} (2016), 075124. % [1605.07199 [cond-mat.mes-hall]].


%@article{PhysRevB.94.075124, title = {Holographic entanglement renormalization of topological insulators},  author = {Wen, Xueda and Cho, Gil Young and Lopes, Pedro L. S. and Gu, Yingfei and Qi, Xiao-Liang and Ryu, Shinsei},  journal = {Phys. Rev. B},  volume = {94},  issue = {7},  pages = {075124},  numpages = {22},  year = {2016},  month = {Aug},  publisher = {American Physical Society},  doi = {10.1103/PhysRevB.94.075124},  url = {https://link.aps.org/doi/10.1103/PhysRevB.94.075124} 


%%%%%%%%%%%%%%%%%%%%%%%
%%%%%%%%%%%%%%%%%%%%%%%%%%%%%%%%%%%%%%%%%%%%%%%%%%%
%%%%%%%%%%%%%%%%%%%%%%%%%%%%%%%%%%%%%%%%%%%%%%%%%%% 

\bibitem{Ryu-Tnayanagi2006}
S.~Ryu and T.~Takayanagi, Holographic Derivation of Entanglement Entropy from the anti--de Sitter Space/Conformal Field Theory Correspondence,   
%\emph{Holographic Derivation of Entanglement Entropy from the anti--de Sitter Space/Conformal Field Theory Correspondence}, 
{\it Phys. Rev. Lett.} \textbf{96} (2006), 181602. % [0605073 [hep-th]].    

%S.~Ryu and T.~Takayanagi,     
%\emph{Holographic Derivation of Entanglement Entropy from the anti--de Sitter Space/Conformal Field Theory Correspondence}, 
%Phys. Rev. Lett\textbf{96} (2006) 181602 [hep-th/0605073].      
%{ @article{PhysRevLett.96.181602,
%  title = {Holographic Derivation of Entanglement Entropy from the anti--de Sitter Space/Conformal Field Theory Correspondence},
%  author = {Ryu, Shinsei and Takayanagi, Tadashi},
%  journal = {Phys. Rev. Lett.},
%  volume = {96},
%  issue = {18},
%  pages = {181602},
%  numpages = {4},
%  year = {2006},
%  month = {May},
%  publisher = {American Physical Society},
%  doi = {10.1103/PhysRevLett.96.181602},
%  url = {https://link.aps.org/doi/10.1103/PhysRevLett.96.181602}
%}  }        
%%%%%%%%%%%%%%%%%%%%%%%%%%%%%%%%%%%%%%%%%%%%%%%%%%%
%%%%%%%%%%%%%%%%%%%%%%%%%%%%%%%%%%%%%%%%%%%%%%%%%%%  
%%%%%%%%%%%%%%%%%%%%%%%%%%%%%%%%%%%%%%%%%%%%%%%%%%%%
\bibitem{Kobayashi-Nishioka-Sato-Watanabe2019}  
N.~Kobayashi, T.~Nishioka, Y.~Sato and K.~Watanabe, Towards a C-theorem in defect CFT, 
{\it JHEP} \textbf{01} (2019), 39. %% [1810.06995 [hep-th]] .
% Towards a C-theorem in defect CFT. J. High Energ. Phys. 2019, 39 (2019)  
  
  
          
%%%%%%%%%%%%%%%%%%%%%%%%%%%%%%%%%%%%%%%%%%%%%%%%%%%
%%%%%%%%%%%%%%%%%%%%%%%%%%%%%%%%%%%%%%%%%%%%%%%%%%%  
%%%%%%%%%%%%%%%%%%%%%%%%%%%%%%%%%%%%%%%%%%%%%%%%%%%%
\bibitem{Ohmori-Tachikawa2015} 
K.~Ohmori and Y.~Tachikawa, Physics at the entangling surface,
{\it J. Stat. Mech.} (2015), P04010. % [1406.4167 [hep-th]].


%%%%%%%%%%%%%%%%%%%%%%%%%%%%%%%%%%%%%%%%%%%%%%%%%%%
%%%%%%%%%%%%%%%%%%%%%%%%%%%%%%%%%%%%%%%%%%%%%%%%%%%  
\bibitem{FLM2013}
T.~Faulkner, A.~Lewkowycz, and J.~Maldacena, Quantum corrections to holographic entanglement entropy, 
{\it JHEP} \textbf{11} (2013), 074. 

%J. High Energ. Phys. 2013, 74 (2013). https://doi.org/10.1007/JHEP11(2013)074 




%%%%%%%%%%%%%%%%%%%%%%%%%%%%%%%%%%%%%%%%%%%%%%%%%%
%%%%%%%%%%%%%%%%%%%%%%%%%%%%%%%%%%%%%%%%%%%%%%%%%%
\bibitem{GKKS2017}
D.~Gaiotto, A.~Kapustin, Z.~Komargodski and N. Seiberg,
Theta, Time Reversal, and Temperature, 
{\it JHEP} \textbf{05} (2017),  091.
%[1703.00501 [hep-th]]. 

%%%%%%%%%%%%%%%%%%%%%%%%%%%%%%%%%%%%%%%%%%%%%%%%%%
%%%%%%%%%%%%%%%%%%%%%%%%%%%%%%%%%%%%%%%%%%%%%%%%%%
\bibitem{Kikuchi-Tanizaki2017}
Y.~Kikuchi and Y.~Tanizaki, Global inconsistency, ’t Hooft anomaly, and level crossing in quantum mechanics,  
{\it Prog. Theor. Exp. Phys.} (2017), 113B05. % [1708.01962 [hep-th]].
%%%%%%%%%%%%%%%%%%%%%%%%%%%%%%%%%%%%%%%%%%%%%%%%%%
%%%%%%%%%%%%%%%%%%%%%%%%%%%%%%%%%%%%%%%%%%%%%%%%%% 

%%%%%%%%%%%%%%%%%%%%%%%%%%%%%
%%%%%%%%%%%%%%%%%%%%%%%%%%%%%
%%%%%%%%%%%%%%%%%%%%%%%%%%%%%%%%%%%%%%%%%%%%%%%%%%%
%%%%%%%%%%%%%%%%%%%%%%%%%%%%%%%%%%%%%%%%%%%%%%%%%%%
%\bibitem{Volovik1991} 
%G.~E.~Volovik, A new class of normal Fermi liquids, 
%JETP Lett. \textbf{53} (1991) 222-225.
%
% A new class of normal Fermi liquids, JETP Lett. \textbf{53} (1991) 222.
%JETP Letters
%1991 vol: 53 (4) pp: 222 - 225
%
%
%%%%%%%%%%%%%%%%%%%%%%%%%%%%%%%%%%%%%%%%%%%%%%%%%%%%%
%%%%%%%%%%%%%%%%%%%%%%%%%%%%%%%%%%%%%%%%%%%%%%%%%%%%
%%%%%%%%%%%%%%%%%%%%%%%%%%%%%%%%%%%%%%%%%%%%%%%%%%%%%
   
\bibitem{Shiozaki-Fujimoto2012}
K. Shiozaki and S. Fujimoto, Green's function method for line defects and gapless modes in topological insulators: Beyond the semiclassical approach,
{\it Phys. Rev.} B\textbf{85} (2012), 085409. % [1111.1685 [cond-mat.mes-hall]]. 
%K. Shiozaki and S. Fujimoto,
%\emph{Green's function method for line defects and gapless modes in topological insulators: Beyond the semiclassical approach,} 
%Phys. Rev. B\textbf{85} (2012) 085409 [1111.1685 [cond-mat.mes-hall]]. 
%https://doi.org/10.1103/PhysRevB.85.085409. 
%Green's function method for line defects and gapless modes in topological insulators: Beyond the semiclassical approach, Phys. Rev. B\textbf{85} (2012) 085409,  https://doi.org/10.1103/PhysRevB.85.085409, arXiv:1111.1685 [cond-mat.mes-hall]


%  @article{PhysRevB.85.085409,
%  title = {Green's function method for line defects and gapless modes in topological insulators: Beyond the semiclassical approach},
%  author = {Shiozaki, Ken and Fujimoto, Satoshi},
%  journal = {Phys. Rev. B},
%  volume = {85},
%  issue = {8},
%  pages = {085409},
%  numpages = {12},
%  year = {2012},
%  month = {Feb},
%  publisher = {American Physical Society},
%  doi = {10.1103/PhysRevB.85.085409},
%  url = {https://link.aps.org/doi/10.1103/PhysRevB.85.085409}
%}



\end{thebibliography}
\end{document}